\documentclass[11pt]{article}

\usepackage{amsmath}
\usepackage{amsthm}
\usepackage{amssymb}
\usepackage{graphicx}
\usepackage{mathrsfs}

\usepackage{algorithmic}
\usepackage{rotating}

\usepackage{hyperref}
\usepackage{breakurl}

\usepackage{dsfont} 

\usepackage[top=1in, bottom=1in, left=1in, right=1in]{geometry}

\usepackage{setspace}
\usepackage{natbib}

\newcommand{\dx}[1]{\ \text{d} #1}
\newcommand{\E}{\mathbb{E}}

\newcommand{\indicator}[1]{\mathds{1}\{ #1 \}}

\newcommand{\er}{Erd\H{o}s-R\'enyi }

\newcommand{\btheta}{\boldsymbol{\theta}}

\newcommand{\A}{\mathbf{A}}

\newcommand{\C}{\mathbf{C}}

\newcommand{\w}{\mathbf{w}}

\newcommand{\bu}{\mathbf{u}}
\newcommand{\bd}{\mathbf{d}}
\newcommand{\Y}{\mathbf{Y}}

\newtheorem{lem}{Lemma}
\newtheorem{defn}{Definition}
\newtheorem{prop}{Proposition}
\newtheorem{cor}{Corollary}

\newtheorem{assumption}{Assumption}

\title{The graphical structure \\ of respondent-driven sampling}

\author{Forrest W. Crawford \\[0.5em]
\small Department of Biostatistics, Yale School of Public Health \\
\small 60 College St. New Haven, CT, USA 06510 \\
\small phone: (203) 785-6125;  fax: (203) 785-6912 \\
\small email: \texttt{forrest.crawford@yale.edu}\\
\small url: \texttt{http://crawford.research.yale.edu} } 

\begin{document}

\maketitle

\begin{abstract} 
\noindent Respondent-driven sampling (RDS) is a chain-referral method for sampling members of a hidden or hard-to-reach population such as sex workers, homeless people, or drug users via their social network. Most methodological work on RDS has focused on inference of population means under the assumption that subjects' network degree determines their probability of being sampled.  Criticism of existing estimators is usually focused on missing data: the underlying network is only partially observed, so it is difficult to determine correct sampling probabilities.  In this paper, we show that data collected in ordinary RDS studies contain information about the structure of the respondents' social network.  We construct a continuous-time model of RDS recruitment that incorporates the time series of recruitment events, the pattern of coupon use, and the network degrees of sampled subjects.  Together, the observed data and the recruitment model place a well-defined probability distribution on the recruitment-induced subgraph of respondents.  We show that this distribution can be interpreted as an exponential random graph model and develop a computationally efficient method for estimating the hidden graph.  We validate the method using simulated data and apply the technique to an RDS study of injection drug users in St. Petersburg, Russia.  \\[1em]
\end{abstract}

\noindent \textbf{Acknowledgements:} This work was supported by NIH Grant KL2 TR000140, NIMH grant P30 MH062294, the Yale Center for Clinical Investigation, and the Yale Center for Interdisciplinary Research on AIDS.  I am especially indebted to Edward H. Kaplan, Robert Heimer, Peter M. Aronow, and Leonid Chindelevitch for providing detailed comments on the manuscript.  
I also thank  
Yakir Berchenko,
Russel Barbour, 
Alexander Bazazi,
Lin Chen,
Krista Gile,
Mark Handcock,
Olga Levina, 
Aleksandr Sirotkin,
Edward White,
Jiacheng Wu,
Alexei Zelenev,
and
Li Zeng
for valuable suggestions and discussion.
The Project 90 data were obtained from the Office of Population Research at Princeton University ({\burl{http://opr.princeton.edu/archive/P90}).
The RDS data presented in the application are from the ``Influences on HIV Prevalence and Service Access among IDUs in Russia and Estonia'' study, funded by NIH/NIDA grant 1R01DA029888 to Robert Heimer and Anneli Uusk\"{u}la (Co-PIs).  
I made use of the Yale University Biomedical High Performance Computing Center, funded by NIH RR029676-01.





\section{Introduction}

Hidden populations such as drug users, men who have sex with men, sex workers, or homeless people are often subject to social stigma or criminalization.  Learning about these populations can be challenging for sociologists, epidemiologists, and public health researchers because potential subjects may fear exposure or prosecution.  Several survey techniques have been developed for sampling from hidden populations, including social link tracing and snowball designs \citep{Goodman1961Snowball,Thompson2000Model}.  Respondent-driven sampling (RDS) is a common survey method for hidden or hard-to-reach populations for which no convenient sampling frame exists \citep{Heckathorn1997Respondent,Broadhead1998Harnessing}.  In RDS, study participants recruit members of their social network who are also members of the hidden population.  Starting with a set of ``seeds'', participants are given a fixed number of coupons tagged with a unique code.  Participants then recruit members of their social network by giving them a coupon.  The recipient of the coupon redeems it at the study site (or over the phone, online, etc.), is interviewed, and receives coupons to recruit others. A dual incentive encourages recruitment: subjects receive a small reward for participating in the study, and for each new subject they recruit. Subjects cannot be recruited more than once, and only a small number of coupons are given to new participants to prevent the local network from being saturated with coupons or emergence of a secondary market for coupons. To safeguard the privacy of subjects not participating in the study, subjects do not reveal the identities of their social contacts to researchers. The only network information typically reported by subjects is their \emph{network degree}, the number of social contacts who are also members of the study population.  

While RDS is an effective procedure for recruiting members of a hidden population, estimation of population characteristics from data obtained by RDS is controversial.  Most methodological work on RDS assumes that the recruitment process takes place on a hidden social network connecting members of the study population.  With the understanding that the structure of this hidden network likely affects individual subjects' likelihood of being recruited, many researchers have sought to determine sampling probabilities for design-based estimation of population means (e.g. HIV prevalence).  \citet{Salganik2004Sampling} construct a model of the recruitment process in which subjects receive only one coupon and can be recruited infinitely many times.  They model the recruitment as a random walk \emph{with replacement} on the hidden population social network.  When this walk is at ``equilibrium'', they argue that the probability that a given subject is sampled is proportional to their network degree.  \citet{Salganik2004Sampling} and \citet{Volz2008Probability} propose a Horvitz-Thompson type estimator for the population mean, where observations are weighted by the inverse of the subject's degree.  \citet{Gile2011Improved} derives a related estimator where sampling is without replacement.

Unfortunately, the characterization of the RDS recruitment process as a sampling design, where sampling probability is a function of network degree alone, suffers from some fundamental flaws.  First, RDS recruitment is always \emph{without replacement}, since subjects cannot be recruited more than once \citep{Heckathorn1997Respondent}; second, a without-replacement random walk on a network is never at equilibrium with respect to its probability of sampling particular subjects -- once a subject is recruited, he or she can never be visited by the recruitment process again; and third, if the recruitment process operates on the social network connecting the sampled individuals and seeds are not chosen at random, the network structure itself determines the probability that a given person will be reached by the recruitment chain.  Indeed, for a given sample size $n$ on a fixed population network, any potential subject whose minimum path length to a seed is greater than $n$ has sampling probability $0$, \emph{regardless of their network degree}.  The random walk characterization of RDS also neglects the fundamental role of coupon depletion in the dynamics of recruitment.  Depletion of certain recruiters' coupons can block paths to isolated parts of the network, providing no way for the recruitment chain to reach some members of the population.  Researchers have raised serious concerns about the empirical properties of population estimates from data obtained by RDS and the \citet{Volz2008Probability} estimator in particular \citep{Goel2010Assessing,Johnston2010Formative,Gile2010Respondent,Salganik2012Commentary,White2012Respondent,Mills2014Errors}.  Studies comparing RDS with traditional sampling or census of the same population have highlighted serious bias in estimates \citep{McCreesh2012Evaluation} or problems with variance estimation \citep{Wejnert2009Empirical}. 

It is difficult to determine the correct sampling probabilities for recruited subjects under RDS because the underlying social network is only partially observed \citep{Gile2010Respondent,Gile2015Network}.  The unobserved links between recruited subjects, and between recruited and unrecruited population members constitute \emph{missing data} in RDS studies. Characterization of the network upon which the sampling process takes place is therefore a major methodological frontier in research on estimation from RDS \citep{Handcock2010Modeling}.  
Perhaps surprisingly, a typical RDS study reveals a great deal of information about the network of respondents: the observed degrees, recruitment chain, and patterns of coupon allocation and depletion are all readily available and provide valuable information about the local structure of the population network.  Insight into the information content of data from RDS studies would clarify exactly which network and population properties researchers can hope to estimate -- and which they cannot -- in real-world studies.  In particular, a better understanding of the network on which RDS recruitment operates could facilitate computation of marginal sampling probabilities similar those calculated by \citet{Gile2015Network} for use in Horvitz-Thompson type estimators for population means \citep[e.g.][]{Volz2008Probability}.  Alternatively, specification of a probability model for dependence between trait values of vertices that share an edge in $G$ may allow regression approaches to population estimation and adjustment for dependence in outcomes induced by the network structure \citep[e.g.][]{Bastos2012Binary}.  An estimate of the sub-network of respondents in an RDS study could also be used to estimate the size of the target population in a manner analogous to the ``network scale-up'' population size estimator \citep{Killworth1998Estimation,Bernard2010Counting,Feehan2014Generalizing}.

In addition to its statistical uses for population-level inference, the sub-network of respondents is of inherent sociological and epidemiological interest.  The network connecting sampled subjects reveals social links between participants and possible avenues for transmission of ideas, behaviors, practices, or infectious agents.  Comprehensive sociometric mapping can be difficult and costly in hidden populations, and many researchers have attempted to estimate epidemiological properties of recruited individuals' networks from recruitment data obtained by RDS \citep[e.g.][]{liu2009egocentric,Cepeda2011Drug,li2011sexual,Stein2014Comparison,Stein2014Online}.  The ability to estimate features of the subgraph of respondents in an RDS study would place sociological and epidemiological inquiries about the local network onto firmer theoretical and methodological ground.  

In other areas of network theory, researchers have made progress in reconstructing networks from partial observation.  When links are missing, some techniques assume that similar subjects are likely to be connected \citep{Leskovec2010Predicting,Lu2011Link,Atchade2011Estimation,Koskinen2013Bayesian}.  When vertices, edges, or egocentric networks are sampled, several authors have proposed ways of estimating global network properties \citep{Smith2012Macrostructure,Bliss2014Estimation,Goyal2014Sampling}, or when vertices can be observed more than once \citep{Yan2013Identifying,Frank1994Estimating}.  Sometimes dynamic or random processes can reveal structural information about networks \citep{Kramer2009Network,Shandilya2011Inferring,Linderman2014Discovering}.  \citet{Gile2011Improved} and \citet{Gile2015Network} present methods for random graph model-assisted inference of the degree distribution from RDS, but still assume sampling probability is a function of network degree alone.

In this paper, we show how to use data from RDS studies to probabilistically reconstruct the social network of respondents.  We first define the observed data under RDS and construct a realistic continuous-time model of the RDS recruitment process on a graph.   The model is a simple and natural formalization of the RDS recruitment procedure initially defined by \citet{Heckathorn1997Respondent}.  Inter-recruitment waiting times carry information about the network edges linking recruiters to unsampled individuals at each moment in time.  We combine this timing information, knowledge of who recruited whom, who had coupons at which times, and the network degrees of recruited subjects to place a well-defined probability distribution on the structure of the recruitment-induced subgraph.  A fundamental result of this paper is that under simple and realistic assumptions, the likelihood of the recruitment process on a hidden graph can be interpreted as exponential random graph model.  We describe a technique for jointly estimating the recruitment-induced subgraph and recruitment rate.  An important feature of the algorithm is a computationally efficient method to calculate the likelihood of the recruitment-induced subgraph.  We validate the proposed technique using simulated and real networks and apply it to an RDS study of injection drug users in St. Petersburg, Russia.  We conclude with a new perspective on the information content of data from RDS studies and ideas for future work.


\section{Preliminaries}

We begin by stating some definitions and assumptions to ensure that the graph inference problem is well-posed (we use the terms ``graph'' and ``network'' interchangeably).  The first is implicit in the foundational work on RDS and guarantees that the objects under study exist \citep{Heckathorn1997Respondent,Salganik2004Sampling,Volz2008Probability}.  
\begin{assumption}
 The hidden population exists and has finite size $N$.  The social network connecting members of the hidden population is an undirected graph $G=(V,E)$ with $|V|=N$ and no parallel edges or self-loops.
 \label{assump:net}
\end{assumption}
\noindent Members of the hidden population are \emph{vertices} in $V$.  A vertex is \emph{recruited} if it is known to the study.  A vertex is a \emph{recruiter} if it has at least one coupon and at least one unrecruited neighbor; a \emph{susceptible vertex} is unrecruited and has at least one neighbor who is a recruiter.  A \emph{susceptible edge} connects a recruiter and a susceptible vertex, and recruitments can only take place across susceptible edges.  A recruited vertex cannot be recruited again.  At the moment it is recruited, a vertex is endowed with a non-negative number of coupons that it may use to recruit its susceptible neighbors.  Every recruitment reduces the number of coupons held by the recruiter by one.  When all the coupons belonging to a recruiter vertex are depleted, the vertex is no longer a recruiter, and any edges incident to it are no longer susceptible.  \emph{Seeds} are recruited vertices chosen from the entire population of vertices by some mechanism, not necessarily random, usually at the beginning of the study. Seeds are not considered to have been recruited by any other vertex.

\begin{defn}[Recruitment-induced subgraph] 
  The recruitment-induced subgraph is $G_S=(V_S, E_S)$, where $V_S\subseteq V$ consists of $n=|V_S|$ sampled vertices (including seeds); and $\{i,j\}\in E_S$ if and only if $i\in V_S$, $j\in V_S$, and $\{i,j\}\in E$.
\end{defn}
 
\begin{defn}[Recruitment graph]
  The directed recruitment graph is $G_R=(V_R,E_R)$, where $V_R=V_S$ is the set of $n$ sampled vertices and a directed edge from $i$ to $j$ indicates that $i$ recruited $j$.
\end{defn}

\noindent Since subjects cannot be recruited more than once, $G_R$ is acyclic.  Assumption \ref{assump:net} does not require that $G$ be connected, nor that the RDS sample take place in the largest connected component, or even a single component.  Therefore the recruitment-induced subgraph $G_S$ may not be connected.  Let $\bd$ be the $n\times 1$ vector of recruited subjects' degrees (in the order of their recruitment into the study) and let $\mathbf{t} = (t_1,\ldots,t_n)$ be the $n\times 1$ vector of recruitment times, where $t_1<\cdots<t_n$.  

\begin{defn}[Coupon matrix]
Let $\C$ be the $n\times n$ coupon matrix whose element $\C_{ij}$ is 1 if subject $i$ has at least one coupon just before the $j$th recruitment event, and zero otherwise.  The rows and columns of $\C$ are ordered by subjects' recruitment time.  
\end{defn}

\noindent The RDS recruitment process reveals only some of this information to researchers. 
\begin{assumption}
  The observed data from an RDS recruitment process consist of $\Y=(G_R,\bd,\mathbf{t},\C)$.  
  \label{assump:obs}
\end{assumption}
\noindent In particular, researchers do not observe the recruitment-induced subgraph $G_S$ of the sampled vertices.  Assumption \ref{assump:obs} is taken from the description of the RDS recruitment procedure given by \citet{Heckathorn1997Respondent}.  The coupon matrix $\C$ is a deterministic function of $G_R$ and the number of coupons given to subjects.  Figure \ref{fig:Y} shows an example graph $G$ and a realization of the RDS recruitment process on $G$.  The recruitment graph $G_R$, recruitment-induced subgraph $G_S$, degree vector $\bd$, recruitment times $\mathbf{t}$, and coupon matrix $\C$ are also shown.

\begin{figure}
\centering
\includegraphics[scale=0.7]{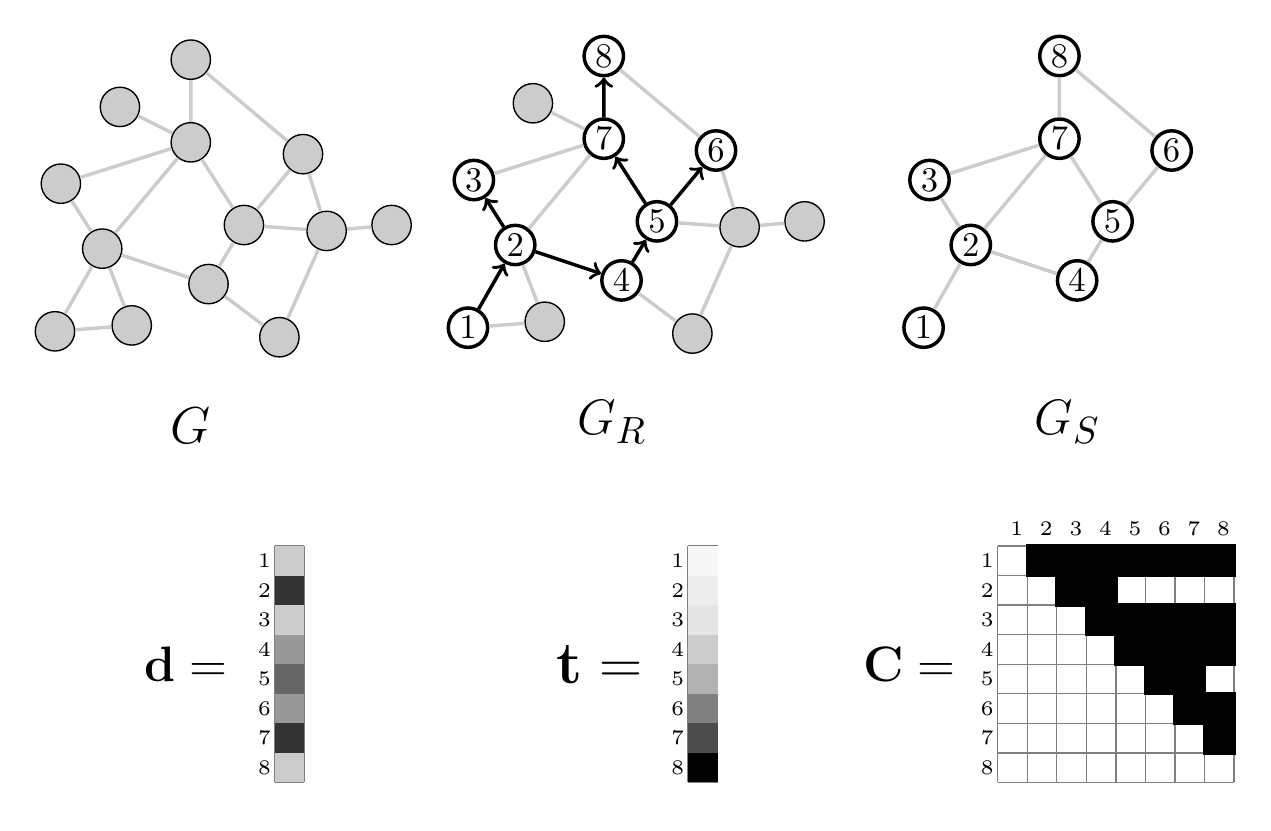}
 \caption{Example of unobserved and observed data in RDS. The true hidden population network is $G$, shown at top left.  One seed is chosen (the vertex marked 1), and the RDS recruitment proceeds with each recruited vertex receiving 2 coupons. The directed recruitment graph $G_R$ is shown superimposed on $G$.  The recruitment-induced subgraph $G_S$ is the subgraph of the recruited vertices. The degrees $\bd=(d_1,\ldots,d_8)$ of each recruited vertex are observed, along with the recruitment times $\mathbf{t}=(t_1,\ldots,t_8)$.  Finally, the coupon matrix $\C$ shows which recruited vertices had at least one coupon just before each recruitment event.  In RDS, researchers observe neither $G$ nor $G_S$; the observed data consist of $\mathbf{Y}=(G_R,\bd,\mathbf{t},\C)$. }
 \label{fig:Y}
\end{figure}

We now state three assumptions about the behavior of recruiters and their knowledge of the recruitment status of their neighbors. 
\begin{assumption}
Vertices become recruiters immediately upon entering the study and receiving one or more coupons.  They remain recruiters until their coupons or susceptible neighbors are depleted, whichever happens first.
\label{assump:recruiter}
\end{assumption}
\begin{assumption}
 When a susceptible neighbor $j$ of a recruiter $i$ is recruited by any recruiter, the edge connecting $i$ and $j$ is immediately no longer susceptible.
 \label{assump:knowing}
\end{assumption}
\noindent By Assumption \ref{assump:knowing}, recruitment is \emph{competitive}: the first recruiter to recruit a given susceptible vertex immediately removes it from the pool of susceptibles.  Finally, we specify a parametric waiting time distribution for the time it takes for a recruiter to recruit a susceptible neighbor.
\begin{assumption}
 The time to recruitment along an edge connecting a recruiter to a susceptible neighbor has exponential distribution with rate $\lambda$, independent of the identity of the recruiter, neighbor, and all other waiting times.
  \label{assump:exp}
\end{assumption}
\noindent By Assumption \ref{assump:exp}, waiting times to recruitment along susceptible edges are independent and elapse concurrently in continuous time, so recruitment is \emph{simultaneous}.  Together, Assumptions \ref{assump:recruiter}, \ref{assump:knowing}, and \ref{assump:exp} place a well-defined probability distribution on the recruitment-induced subgraph of respondents.

\subsection{Consequences of the waiting time assumption}

The following results follow directly from Assumption \ref{assump:exp}.  Let $R$ be the set of recruiters with coupons and let $S$ be the set of susceptible vertices at a certain moment in the recruitment process.  Let $S_u$ be the set of susceptible vertices that are neighbors of the recruiter $u\in R$.  Likewise, let $R_v$ be the set of possible recruiters of a susceptible vertex $v\in S$.  Clearly, $v\in S_u$ if and only if $u\in R_v$.  
\begin{prop}
Given that the recruiter $u$ recruits one of its susceptible neighbors $v\in S_u$ before any other recruiter, the waiting time to this recruitment event is distributed as $\text{Exponential}\left(\lambda |S_u|\right)$.  The probability that the susceptible vertex $v\in S_u$ is the next recruit is uniform $1/|S_u|$, independent of the waiting time to the recruitment event.
\label{prop:unif}
\end{prop}
\begin{prop}
  The waiting time to the next recruitment of any susceptible vertex is distributed as 
    $\text{Exponential}\left(\lambda\sum_{u\in R} |S_u|\right)$ .
The probability that the susceptible vertex $v\in S$ is the next recruit is 
\begin{equation}
  \frac{|R_v|}{\sum_{k\in S} |R_k| },
  \label{eq:myrecprob}
\end{equation}
independent of the waiting time.
\label{prop:wt}
\end{prop}
\noindent Proofs of Propositions \ref{prop:unif} and \ref{prop:wt} are given in the Appendix. Intuitively, Proposition \ref{prop:wt} means that the new recruited vertex is chosen with probability proportional to the number of edges along which it can be recruited.  These results formalize the consequences of simultaneous and competitive recruitment in continuous time.

Interestingly, Assumptions 3-5 and the resulting recruitment probability \eqref{eq:myrecprob} differ starkly from the recruitment dynamics used in simulations by other researchers to test the performance of estimators for RDS.  \citet{Gile2010Respondent} simulate the RDS recruitment process by first choosing seeds, after which ``[s]ubsequent sample waves were selected without-replacement by sampling up to two nodes at random from among the unsampled alters of each sampled node''.  This leads us to a brief Corollary establishing the difference between these approaches.
\begin{cor}
Assumptions 3-5 (simultaneous and competitive recruitment) result in different recruitment probabilities than the RDS recruitment implementation of \citet{Gile2010Respondent}.
\label{cor:recprob}
\end{cor}
\noindent A proof is given in the Appendix.  Figure \ref{fig:recprob} shows an example in which the probability of the next recruited vertex under the model developed in this paper is different from that proposed by \citet{Gile2010Respondent}.  The process defined by \citet{Gile2010Respondent} requires that recruiters ``take turns''.  This approach implicitly requires that recruiters have knowledge about the behavior of other recruiters -- even those to whom they are not connected in the network. 
This process induces a different distribution on the susceptible degree, and hence overall degree, of the new recruit than the model described in Assumptions 3-5 of this paper.  Most existing methods for population inference from RDS data depend intimately on the degree distribution of recruited vertices \citep[e.g.][]{Salganik2004Sampling,Volz2008Probability,Gile2011Improved}, so it is important to highlight scenarios when methods for simulation of recruitment dynamics differ.

\begin{figure}
  \centering
  \includegraphics[scale=0.7]{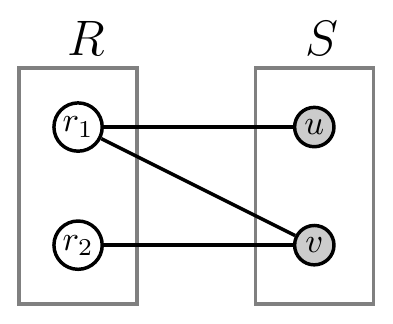}
 \caption{Probability of the next recruited vertex.  Consider a set $R=\{r_1,r_2\}$ of two recruiters and a set $S=\{u,v\}$ of two susceptible vertices $r_1$ is connected to both $u$ and $v$ and $r_2$ is connected only to $v$. Under simultaneous and competitive recruitment (Assumption \ref{assump:exp}), the probability that a given susceptible vertex is recruited next is proportional to the number of susceptible edges incident to it. Then $u$ is the next recruit with probability $1/3$ and $v$ with probability $2/3$.  \citet{Gile2010Respondent} simulate the RDS recruitment process with $u\in R$ chosen first with probability $1/2$, followed by $v\in S_u$. This procedure results in recruitment probabilities $1/4$ and $3/4$ for $u$ and $v$ respectively.}
 \label{fig:recprob}
\end{figure}


\section{Likelihood of the recruitment time series}
\label{sec:lik}

Proposition \ref{prop:wt} shows that under Assumptions \ref{assump:recruiter}-\ref{assump:exp}, the rate of recruitment is proportional to the number of susceptible edges.  Given a realization of the recruitment-induced subgraph $G_S$, it is not immediately obvious how to quickly determine the number of susceptible edges just before each recruitment.  When a given susceptible vertex is recruited, all susceptible edges incident to it disappear from the set of susceptible edges (Assumption \ref{assump:knowing}).  Furthermore, the newly recruited vertex now has coupons, so there may be new susceptible edges connected to it.  Finally, if the new vertex is not a seed, its recruiter has used one coupon; if its coupons are now depleted, any other susceptible edges incident to the recruiter are no longer susceptible.  Clearly the number of susceptible edges can increase, decrease, or stay the same from one recruitment to the next.  In this Section, we derive a computationally efficient representation of the likelihood of the recruitment time series using matrix algebra.  This approach obviates costly enumeration of all $|E_S|$ edges to determine whether they are susceptible at each step in the recruitment process.  A preliminary definition will assist us in this task.  Let $\indicator{X}$ be the indicator of an event $X$, which takes value 1 when $X$ is true, and zero otherwise.
\begin{defn}[Compatibility] 
  An estimated subgraph $\widehat{G}_S=(\widehat{V}_S,\widehat{E}_S)$ is compatible with the observed data if the following conditions are met: 
  \begin{enumerate}
    \item the vertices in the estimated subgraph are identical to the set of recruited vertices: $v\in\widehat{V}_S$ if and only if $v\in V_R$; 
    \item all directed recruitment edges are represented as undirected edges: if $(i,j)\in E_R$ then $\{i,j\}\in\widehat{E}_S$; 
    \item the number of edges in $G_S$ belonging to each sampled vertex does not exceed the vertex's degree: for all $v\in V_R$, $\sum_{u\in V_R\setminus v} \indicator{\{u,v\}\in \widehat{E}_S} \leq d_v$, where $d_v$ is the degree of vertex $v$.  
  \end{enumerate}
  \label{def:compatibility}
\end{defn}
  \noindent These compatibility conditions provide topological constraints on the structure of $G_S$.  Combining these with the likelihood of the recruitment time series places a probability distribution on the topology of $G_S$.

\begin{figure}
 \centering
 \includegraphics[scale=0.7]{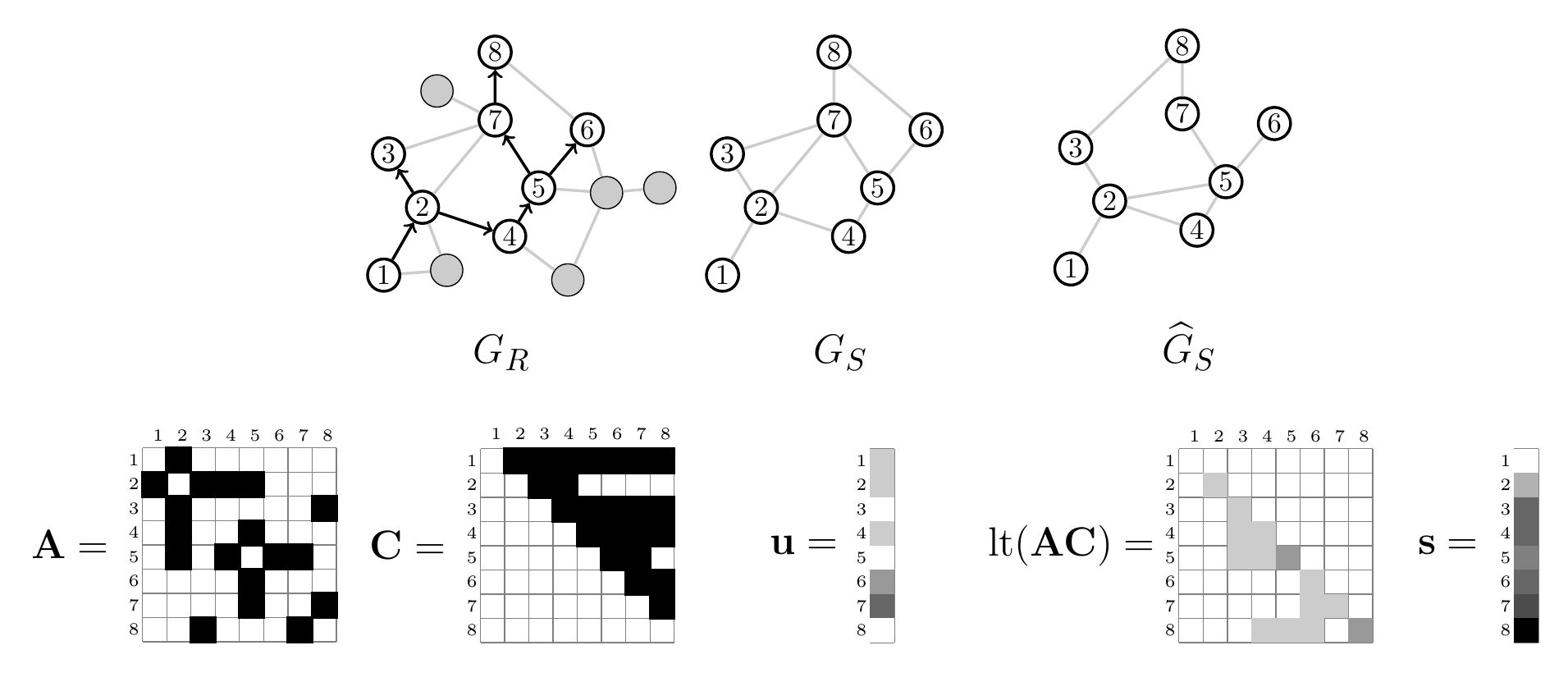}
 \caption{Examples of matrices used to calculate the recruitment time series likelihood.  At top left is the recruitment graph $G_R$ overlaid on the population graph $G$, with recruited vertices numbered and other vertices and edges in gray. 
   The true recruitment-induced subgraph $G_S$ is not directly observed.  We estimate $G_S$ by $\widehat{G}_S$ and let $\mathbf{A}$ be the adjacency matrix of $\widehat{G}_S$.  The coupon matrix $\C$ and the number of pendant edges attached to each recruited vertex $\bu$ are shown.  Pendant edges connect recruited vertices to unknown/unsampled vertices.  The $i,j$th element of $\text{lt}(\A\C)$ is the number of recruiters connected to $i$ just before the $j$th recruitment event.  }
 \label{fig:example}
\end{figure}

Let $\A$ be the $n\times n$ adjacency matrix (sociomatrix) of a compatible estimate $G_S$, where the rows and columns of $\A$ correspond to subjects in the order of their recruitment.  The product of $\A$ and the coupon matrix $\C$ gives an $n\times n$ matrix whose elements describe the number of recruiters connected to each vertex in $G_S$ over time.  Let $\w=(0,t_2-t_1,\ldots,t_n-t_{n-1})$ be the $n\times 1$ vector of waiting times between recruitments.  Let $\bu$ be the $n\times 1$ vector of the number of edge ends belonging to each vertex (in the order of recruitment) that are not connected to any other sampled vertex.  With this notation, we see that vertex degree is \emph{conserved}, $\bu_i = \bd_i  - \sum_j \A_{ij}$.  When $j\leq i$, $\{\A\C\}_{ij}$ is the number of recruiters connected to $i$ just before the time $t_j$ of the $j$th recruitment.  Then $\text{lt}(\A\C)$, the lower triangle of $\A\C$, is the number of recruiters connected to each vertex at each time before recruitment of that vertex.  Likewise, the $j$th element of $\C'\bu$ is the number of susceptible edges connecting sampled vertices to unsampled vertices at time $t_j$.  Figure \ref{fig:example} shows examples of these matrices.  Finally, let $M$ be the set of seeds.  The following Proposition gives the likelihood of the recruitment time series. 
\begin{prop}
  Under Assumptions 1-5, the likelihood of the recruitment time series is
  \begin{equation}
    L(\mathbf{w}|G_S,\lambda) = \left(\prod_{k\notin M} \lambda \mathbf{s}_k\right) \exp\left[ -\lambda \mathbf{s}'\w \right]
\label{eq:lik}
\end{equation}
where 
\begin{equation}
\mathbf{s} = \text{lt}(\A\C)'\mathbf{1} + \C'\bu 
  \label{eq:s}
\end{equation}
is a vector whose elements are the number of susceptible edges just before each recruitment event.
\label{prop:lik}
\end{prop}
\noindent A proof is given in the Appendix.  As before, the rate of recruitment is proportional to the number of susceptible edges, and Proposition \ref{prop:lik} generalizes Proposition \ref{prop:wt} by providing an explicit expression for the number of susceptible edges at each step, and allowing for seeds to be added at any time.

\subsection{The likelihood is an exponential random graph model} 

While \eqref{eq:lik} is the likelihood of the recruitment time series $\w$, we can also view it as a function of the recruitment-induced subgraph adjacency matrix $\A$ with $\w$ held fixed.  Consider the statistic $T(\A)=-\lambda \mathbf{s}$, where $\mathbf{s}$ is defined by \eqref{eq:s}, $\btheta=\mathbf{w}$, and $B(\mathbf{A})=\sum_{k\notin M} \log(\lambda \mathbf{s}_k)$.  Then we can re-normalize the likelihood \eqref{eq:lik} to form the probability
\begin{equation}
  \Pr(\A|\btheta) = \frac{\exp[ T(\A)'\btheta + B(\A)]}{\kappa(\btheta)} 
\label{eq:ergm}
\end{equation}
where $\kappa(\btheta)$ is a normalizing constant that does not depend on $\A$.  It is clear from \eqref{eq:ergm} that $\Pr(\A|\btheta)$ is a member of the exponential family of distributions.  In particular, it is an exponential random graph model (ERGM), also known as a $p^*$ graph \citep{Frank1986Markov,Wasserman1996Logit}.  

Regardless of whether we view \eqref{eq:lik} as the likelihood of the random waiting times $\w$ or as the probability of the random graph $G_S$, the inference procedure we develop below benefits from Markov chain Monte Carlo algorithms developed for sampling edges in ERGMs \citep[see][for example]{Snijders2002Conditional}.  One of the most important properties of ERGMs is that it is easy to sample $\mathbf{A}$ for a given $\btheta$.  To illustrate, consider the adjacency matrices of two compatible graphs, $\mathbf{A}_1$ and $\mathbf{A}_2$.  The ratio of probabilities, used below in formulation of the Metropolis-Hastings acceptance probability, is
\begin{equation}
  \frac{\Pr(\mathbf{A}_1|\btheta)}{\Pr(\mathbf{A}_2|\btheta)} = \exp\left[ \big( T(\mathbf{A}_1) - T(\mathbf{A}_2) \big)'\btheta + B(\mathbf{A}_1) - B(\mathbf{A}_2)\right]. 
\end{equation}
As we show below, the \emph{change statistic} in the exponential term is straightforward to calculate without the matrix multiplication seemingly required by \eqref{eq:s}.


\section{Estimating $G_S$ and $\lambda$}

Together, the compatibility conditions (Definition \ref{def:compatibility}) and Proposition \ref{prop:lik} make possible simultaneous estimation of the recruitment-induced subgraph $G_S$ and the waiting time parameter $\lambda$ under the recruitment model.  We define the joint posterior distribution as
\begin{equation}
  p(G_S,\lambda|\Y) \propto L(\w|G_S,\lambda)\ \Pr(G_S)\ \pi(\lambda)\ 
  \label{eq:post}
\end{equation}
where $\Y=(G_R,\mathbf{C},\bd,\mathbf{t})$ is the observed data, and $\Pr(G_S)$ and $\pi(\lambda)$ are prior distributions.  In the rest of this paper, we take the uniform distribution over compatible graphs, $\Pr(G_S)\propto 1$, but in principle $\Pr(G_S)$ could be any distribution over $\mathcal{C}(G_R,\bd)$.

\subsection{Adding and removing edges}

To estimate $G_S$, it is helpful to have a recipe for generating compatible subgraphs.  Let $\mathcal{C}(G_R,\bd)$ denote the set of all recruitment-induced subgraphs that are compatible with the observed data $G_R$ and $\bd$.  To obtain a new compatible subgraph $\widehat{G}_S$ from a current compatible subgraph $G_S=(V_S,E_S)$, we randomly choose two vertices $i$ and $j$, where $i\neq j$.  If $\{i,j\}\notin E_S$, $\bu_i>0$, and $\bu_j>0$, then we propose to add the edge $\{i,j\}$ to $E_S$.  Alternatively, if $\{i,j\}\in E_S$ and $\{i,j\}\notin E_R$, then we propose to remove the edge $\{i,j\}$ from $E_S$.  If neither of these conditions hold, we pick another pair $\{i,j\}$ and try again.  This procedure is described formally in the following algorithm.
\begin{algorithmic}[1]
  \LOOP
  \STATE Choose two vertices $i$ and $j$ at random, with $i<j$.
  \IF{ $\{i,j\}\notin E_S$ and $\bu_i\geq 1$ and $\bu_j\geq 1$ }
      \STATE let $E_S^+ = \{i,j\} \cup E_S$ and $G_S^+=(V_S,E_S^+)$ 
      \STATE let $\bu_k^+=\bu_k$ for all $k\neq i,j$ and $\bu_i^+=\bu_i-1$, $\bu_j^+=\bu_j-1$. 
      \RETURN $G_S^+$ and $\bu^+$
  \ELSIF{ $\{i,j\}\in E_S$ and $\{i,j\}\notin E_R$ }
       \STATE let $E_S^- = E_S\setminus \{i,j\}$ and $G_S^-=(V_S,E_S^-)$  
       \STATE $\bu_i^-=\bu_i+1$ and $\bu_j^-=\bu_j+1$
      \RETURN $G_S^-$ and $\bu^-$
      \ENDIF
   \ENDLOOP 
\end{algorithmic}
This procedure chooses a vertex pair $\{i,j\}$ uniformly at random from all pairs whose change (addition or removal of the edge between $i$ and $j$) would result in a compatible graph.  The space of compatible subgraphs $\mathcal{C}(G_R,\bd)$ is connected via proposals of this type.  To see why this is so, consider two compatible graphs $G_S^1$ and $G_S^2$ in $\mathcal{C}(G_R,\bd)$.  Let $G_R^r=(V_S,E_R^r)$ be the undirected recruitment graph obtained by making each edge in the directed recruitment graph $G_R$ reciprocal.  By definition, $G_R^r$ is a subgraph of every graph in $\mathcal{C}(G_R,\bd)$.  From $G_S^1$, we can obtain $G_R^r$ by successively removing non-recruitment edges, one at a time and each of these steps occurs with positive probability.  From $G_R^r$ we can obtain $G_S^2$ by adding non-recruitment edges, one at a time.  Since we can reach $G_S^2$ from $G_S^1$ in a similar manner, $\mathcal{C}(G_R,\mathbf{d})$ is connected via the given proposal algorithm.

\subsection{Monte Carlo sampling}

\label{sec:mc}

To draw samples $(G_S,\lambda)$ from the posterior distribution $p(G_S,\lambda|\Y)$, we describe a Metropolis-within-Gibbs sampling scheme. This involves first sampling $G_S$ conditional on $\lambda$, then sampling $\lambda$ conditional on $G_S$. By alternating these steps, we define a reversible Markov chain whose equilibrium distribution is the given by \eqref{eq:post}. To sample $G_S$ conditional on $\lambda$, suppose $\lambda$ is fixed and we have a compatible subgraph $G_S$. We generate a new compatible subgraph $G_S^*=(V_S,E_S^*)$ using the algorithm given above. We take the uniform prior distribution over the recruitment-induced subgraph: $\Pr(G_S) = 1/|\mathcal{C}(G_R,\bd)|$ for every $G_S\in \mathcal{C}(G_R,\bd)$.  To decide whether to accept a proposal $G_S^*$ as a sample from the conditional distribution $p(G_S|\lambda,\Y)$, we form the Metropolis-Hastings acceptance probability
\begin{equation}
    \rho = \min\left\{ 1, \frac{L(\w|G_S^*,\lambda)}{L(\w|G_S,\lambda)} \cdot \frac{\Pr(G_S|G_S^*)}{\Pr(G_S^*|G_S)} \right\}, 
     \label{eq:mhGS}
\end{equation}
and we accept the proposed graph $G_S^*$ with probability $\rho$.  Section \ref{sec:reclr} gives a simple and computationally efficient expression for the likelihood ratio, and Section \ref{sec:propratio} gives a derivation of the proposal ratio $\Pr(G_S|G_S^*)/\Pr(G_S^*|G_S)$.

To sample $\lambda$ conditional on $G_S$, we employ a Metropolis-Hastings step based on an approximation to the conditional distribution of $\lambda$.  From \eqref{eq:lik}, we can easily find the maximum likelihood estimator of $\lambda$, 
\begin{equation}
  \hat\lambda = \frac{n-|M|}{\mathbf{s}'\mathbf{w}} 
  \label{eq:lamhat}
\end{equation}
with asymptotic variance $\sigma^2 = \lambda^2/(n-|M|)$.  Let 
\begin{equation}
  g(\lambda|G_S) = \frac{1}{\sqrt{2\pi}\sigma} \exp[-(\lambda-\hat\lambda)^2/\sigma^2 ]
\end{equation}
be a normal approximation to the conditional distribution of $\lambda$ given $G_S$.  Suppose $\lambda$ is the current value and we propose a new value $\lambda^*$ from $g(\lambda|G_S)$.  We accept the proposal with probability 
\begin{equation}
\rho = \text{min}\left\{\frac{L(\w|G_S,\lambda^*)\ \pi(\lambda^*)}{L(\w|G_S,\lambda)\ \pi(\lambda)} \cdot \frac{g(\lambda|G_S)}{g(\lambda^*|G_S)} \right\} .
\label{eq:mhlam}
\end{equation}

\subsection{Computing the likelihood ratio}

\label{sec:reclr}

The ratio of likelihoods in \eqref{eq:mhGS} can be computed very efficiently.  Since by definition vertex $i$ was recruited before $j$, we have $t_i<t_j$.  Let $t_i^*$ be the minimum of the time that vertex $i$ used its last coupon and $t_n$, the end of the study.  Let $\mathbf{s}$ be given by \eqref{eq:s} for a particular realization of $G_S$ and let $\mathbf{s}^+$ be the susceptible vector after addition of an edge between $i$ and $j$, where $i<j$.  Likewise let $\mathbf{s}^-$ be the susceptible vector after removal of an edge between $i$ and $j$.  The following result will be useful in computing the likelihood ratio in a simple way.
\begin{lem}
  Given $\mathbf{s}$, $i$, and $j$, where $t_i<t_j$, the vectors $\mathbf{s}^+$ and $\mathbf{s}^-$ are given by
\begin{equation}
  \mathbf{s}_k^+ = \mathbf{s}_k - \indicator{k>j}C_{ik} - C_{jk}
  \label{eq:skadd}
\end{equation}
\begin{equation}
  \mathbf{s}_k^- = \mathbf{s}_k + \indicator{k>j}C_{ik} + C_{jk}
  \label{eq:skrem}
\end{equation}
for $k=1,\ldots,n$.  
\label{lem:updates}
\end{lem}
\noindent A proof of Lemma \ref{lem:updates} is given in the Appendix.  The following Proposition establishes likelihood ratios for addition and removal of edges in $G_S$.
\begin{prop}
Suppose $G_S=(V_S,E_S)$ has $\{i,j\}\notin E_S$, $\bu_i\geq 1$, and $\bu_j\geq 1$.  For a proposal $G_S^+=(V_S,E_S^+)$ identical to $G_S$ except that $\{i,j\}\in E_S^+$, 
the likelihood ratio is
\begin{equation}
  \frac{L(\w|G_S^+,\lambda)}{L(\w|G_S,\lambda)} = \left(\prod_{k\notin M} \frac{\mathbf{s}_k^+}{\mathbf{s}_k} \right) e^{\lambda (t_i^* - \min\{t_j,t_i^*\} + t_j^* - t_j )}.
  \label{eq:addratio}
\end{equation}
Now suppose $G_S=(V_S,E_S)$ has $\{i,j\}\in E_S$ and $\{i,j\}\notin E_R$.  For a proposal $G_S^-=(V_S,E_S^-)$ identical to $G_S$ except that $\{i,j\}\notin E_S^-$, 
the likelihood ratio is
\begin{equation}
  \frac{L(\w|G_S^-,\lambda)}{L(\w|G_S,\lambda)} = \left(\prod_{k\notin M} \frac{\mathbf{s}_k^-}{\mathbf{s}_k}\right) e^{-\lambda (t_i^* - \min\{t_j,t_i^*\} + t_j^* - t_j )} .
  \label{eq:remratio}
\end{equation}
\label{prop:ratio}
\end{prop}
\noindent A proof of Proposition \ref{prop:ratio} is given in the Appendix.  These expressions depend only on a simple \emph{change statistic} and do not require evaluation of the matrix products required by the likelihood \eqref{eq:lik} for the proposal.  Indeed, the change in susceptible edge time, $t_i^*-\min\{t_j,t_i^*\}+t_j^*-t_j$, does not depend on any unknown parameters and can be computed in advance for every $i$ and $j>i$.  Likewise, $\mathbf{s}_k$ can be updated at each step using \eqref{eq:skadd} and \eqref{eq:skrem} without computing the matrix product in \eqref{eq:s}. More generally, the computational burden of the procedure scales with the sample size $n$, and is not affected by the total size $N=|V|$ of the target population.

\subsubsection{Proposal ratio}

\label{sec:propratio}

To define the subgraph proposal ratio in \eqref{eq:mhGS}, consider a given subgraph $G_S$.  The number of possible vertex pairs between which an edge can be added is
\begin{equation}
  \text{Add}(G_S) =  \sum_{i=1}^{n-1} \sum_{j=i+1}^n \indicator{
    \{i,j\}\notin E_S \text{ and } \bu_i\geq 1 \text{ and } \bu_j\geq 1 } .
\label{eq:add}
\end{equation}
Likewise, for a proposed removal of an edge, the number of possible vertex pairs is
\begin{equation}
  \text{Remove}(G_S) =  \sum_{i=1}^{n-1} \sum_{j=i+1}^n \indicator{
    \{i,j\}\in E_S \text{ and }
    \{i,j\}\notin E_R } .
    \label{eq:remove}
\end{equation}
Then the proposal probability for obtaining $G_S^*$ from $G_S$ is 
\begin{equation}
\Pr(G_S^*|G_S) = 1/(\text{Add}(G_S) + \text{Remove}(G_S)) .
\end{equation}
and the ratio of backward and forward proposal probabilities for an addition is
\begin{equation}
  \frac{\Pr(G_S|G_S^*)}{\Pr(G_S^*|G_S)} = 
\frac{1/(\text{Add}(G_S^*) + \text{Remove}(G_S^*)) }{1/(\text{Add}(G_S) + \text{Remove}(G_S)) } = 
\frac{\text{Add}(G_S) + \text{Remove}(G_S)}{\text{Add}(G_S^*) + \text{Remove}(G_S^*)}
\label{eq:naddratio}
\end{equation}

\subsection{Maximum likelihood estimation}

\label{sec:ml}

In Section \ref{sec:mc} we described a Markov chain Monte Carlo algorithm for drawing from the posterior distribution of $G_S$.  However, this approach can be time consuming, and may be unnecessary if only a single ``most-likely'' subgraph $G_S$ is desired.  The same routine can be used to find a maximum likelihood (or maximum \emph{a posteriori}) estimate of $G_S$ and $\lambda$.  This Monte Carlo optimization approach is called ``simulated annealing'' \citep[see, e.g.][for details]{Robert2004Monte}.  Let $\gamma>0$ be a scale factor and let 
\begin{equation}
 L_\gamma(\w|G_S,\lambda) = \exp\left[ -\left(\lambda \mathbf{s}'\w + \sum_{k\notin M}\log(\lambda \mathbf{s}_k)\right)/\gamma \right]
\end{equation}
Define a sequence $\gamma_1,\gamma_2,\ldots$ such that $\lim_{i\to\infty}\gamma_i=0$. At each iteration $i$, we propose $G_S^*$ and compute \eqref{eq:mhGS} with $L_{\gamma_i}(\w|G_S^*,\lambda)$ to accept the proposal with probability $\rho$.  Once $G_S$ is sampled, the most likely $\lambda$ is obtained by \eqref{eq:lamhat}.  The joint sequence of sampled subgraphs and $\lambda$'s tends toward the maximum likelihood estimates very rapidly.


\section{Validation using simulations}

\label{sec:sim}

In simulation studies, reasonably accurate reconstruction of the recruitment-induced subgraph $G_S$ can be achieved using the proposed recruitment model \eqref{eq:lik}.  We analyze the performance of reconstruction in simulated networks and a real-world social network.  Conditional on the population network, we simulate the RDS recruitment process with $n$ subjects, $|M|$ seeds, and recruitment rate $\lambda$, under Assumptions \ref{assump:recruiter}-\ref{assump:exp}.  From the simulated recruitment data, we extract the observed data $\Y=(G_R,t,\bd,\C)$ in accordance with Assumption \ref{assump:obs}.  We place a Gamma prior distribution on the waiting time parameter, $\pi(\lambda) \propto \lambda^{\eta-1}e^{-\xi\lambda}$ where $\eta>0$ and $\xi>0$.  We assess the accuracy of reconstruction over 100 repetitions of simulated RDS recruitment over different networks, and for each simulated dataset we find the joint maximum \emph{a posteriori} (MAP) estimate of $G_S$ and $\lambda$ using the procedure outlined in Section \ref{sec:ml}.  MAP estimates represent the mode of the posterior distribution over $(G_S,\lambda)$ and provide a convenient point estimate for comparing results over many repetitions of the simulation.  We analyze the performance of reconstruction on simulated \er networks and a real-world network derived from a network study heterosexuals at high risk of contracting HIV in Colorado Springs, CO, USA from 1988-1990 called Project 90 \citep{woodhouse1994mapping,klovdahl1994social,Rothenberg1995Social,Potterat2004Network}.  We also evaluate reconstruction under mis-specification of the waiting time model, in which Assumption \ref{assump:exp} is violated. Reconstruction remains robust, with corresponding bias in estimates of $\lambda$. We present detailed simulation results in the online Appendix.

A simple illustrative example of MAP estimation with $\lambda=1$, $n=50$, $|M|=1$ seed, and three coupons per subject is shown in Figure \ref{fig:simexample}.  The prior distribution of $\lambda$ has mean 1 and SD 0.1. The population network is derived from the Project 90 network data, described in detail below. The top row shows the true subgraph $G_S$ with the recruitment graph $G_R$ overlaid (arrows indicate recruitment edges), adjacency matrices of $G_R$, $G_S$, and an estimate $\widehat{G}_S$.  The bottom row shows the number of edges $|\widehat{E}_S|$ in the estimated subgraph at each iteration, the trace of $\lambda$, log posterior values, and accuracy.

\begin{sidewaysfigure}
  \begin{center}
    \includegraphics[width=0.7\textwidth]{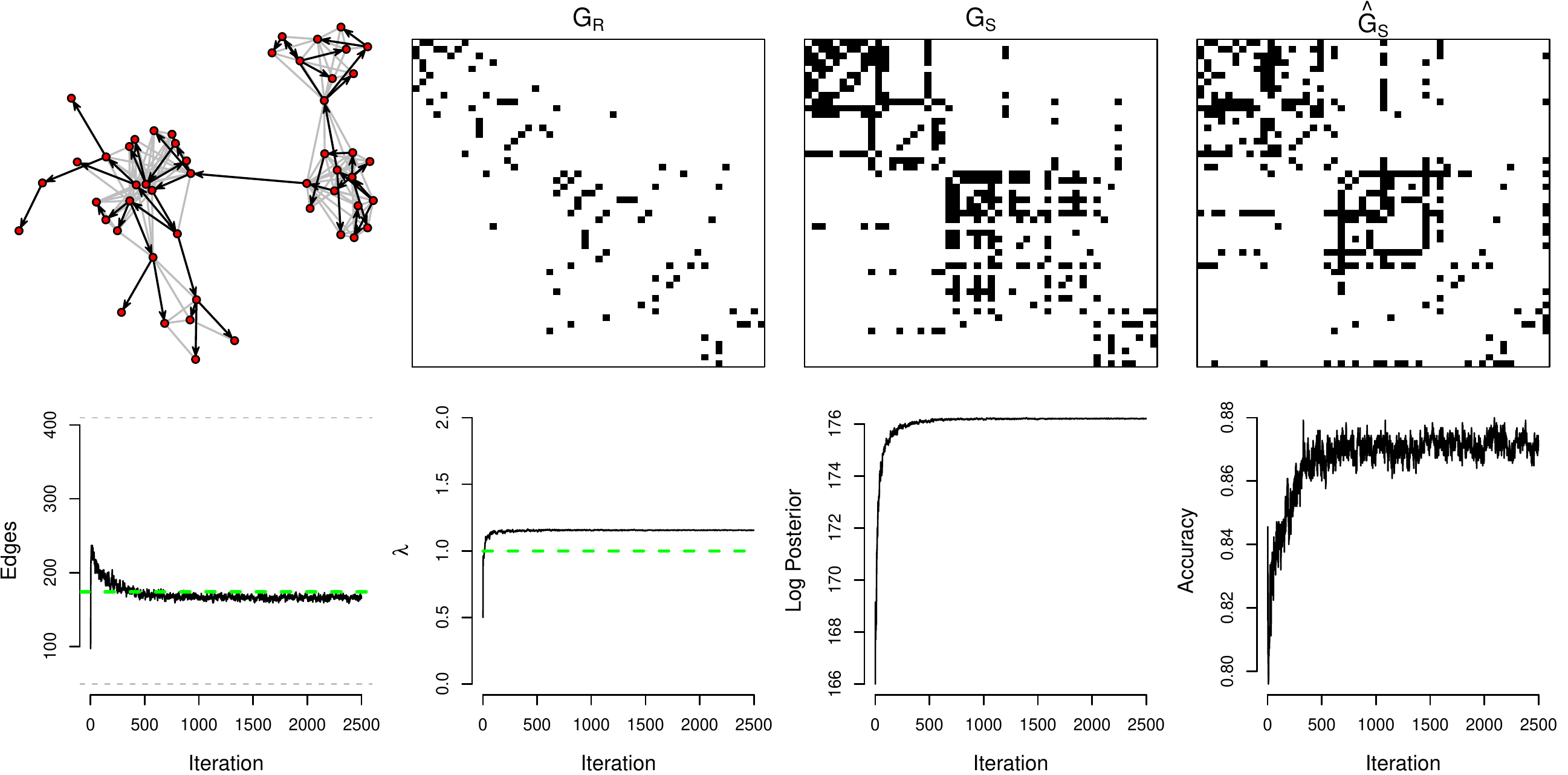}
  \end{center}
  \caption{Example maximum \emph{a posteriori} reconstruction of $G_S$ and estimation of $\lambda$ from the Project 90 network and simulated RDS with $\lambda=1$, $n=50$, $|M|=1$ seed, and three coupons per subject. The Gamma prior for $\lambda$ has mean 1 and SD 0.1.  The top row shows the recruitment-induced subgraph $G_S$ with the directed recruitment graph $G_R$ overlaid, and the adjacency matrices of $G_R$, $G_S$, and a sample reconstruction $\widehat{G}_S$ of $G_S$. The bottom row shows the traces of the estimated number of edges $|E_S|$, $\lambda$, the unnormalized log-posterior value, and the accuracy of $\widehat{G}_S$ as a predictor of $G_S$ at each iteration.  Accuracy is defined as the number of correct entries in the adjacency matrix of $G_S$ divided by the total number of entries. The true number of edges is $|E_S|=174$ and the number of edges in the estimated graph is 169.  The estimated $\lambda$ is 1.15. }
  \label{fig:simexample}
\end{sidewaysfigure}


\section{Application}

The HIV epidemic in St. Petersburg, Russia is concentrated in people who inject drugs (PWID). At least 12,000 people are registered as drug users, but the number of current PWID is likely much higher \citep{Heimer2010Estimation}. Injection drug use is highly stigmatized in the Russian Federation, and criminal penalties for drug possession can be severe.  PWID suffer from high rates of HIV infection and may lack access to treatment and health-related educational resources \citep{Niccolai2010High,Niccolai2011Estimates}.  

\subsection{Study description}

\begin{figure}
  \begin{center}
    \includegraphics[width=0.7\textwidth]{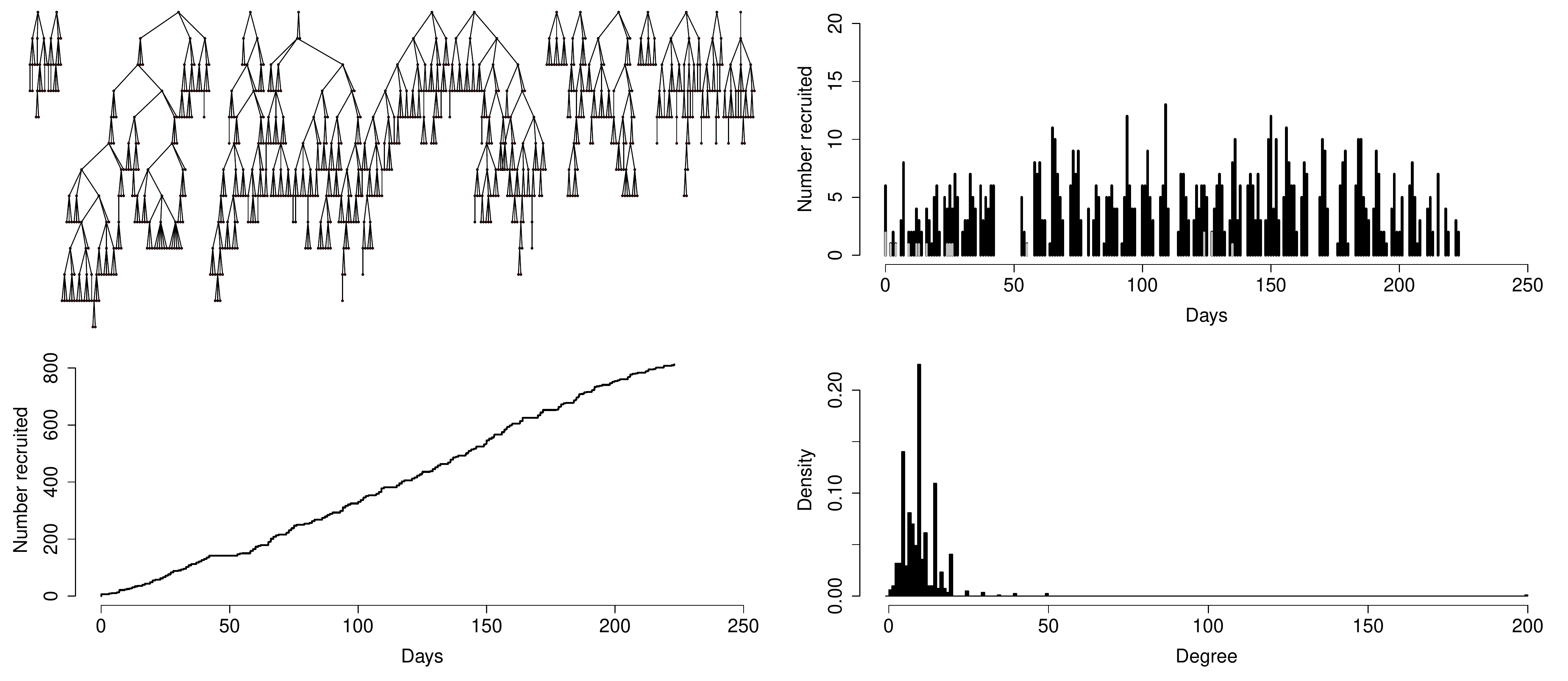}
  \end{center}
  \caption{Raw data from an RDS sample of $n=813$ people who inject drugs in St. Petersburg, Russia.  In the top left panel, fourteen RDS recruitment chains originating from different seeds are shown.  Recruited subjects are organized into ``waves'' along the vertical axis.  The top right panel shows the number of subjects interviewed on each day of the study, with seeds indicated by gray bars.  The bottom left panel shows the cumulative number of recruits over the course of the study, and the bottom right panel shows a histogram of the reported degrees of subjects, with bin size one.}
  \label{fig:rawdata}
\end{figure}

As part of a study to assess perceived barriers to use of HIV prevention and treatment services, $n=813$ PWID were recruited using RDS in St. Petersburg during 2012-2013.  Outreach workers identified 17 seed subjects using venue-based sampling in six city districts.  Interviews collected demographic information, injection practices, sex practices, mental health measures, and knowledge of HIV/AIDS and tuberculosis resources, but we focus solely on network structure in this analysis.  Figure \ref{fig:rawdata} shows the raw RDS data: the recruitment trees, number of new recruits per day, cumulative number of recruits, and reported network degrees.

Participation in the study was limited to current injection drug users over the age of 21 who had injected within the previous four weeks.  Subjects' status as PWID was verified either by inspection of arms for injection marks or explanation of drug preparation.  Subjects received a voucher with a value of about US\$20 for being interviewed, and a secondary reward with value about US\$10 for recruiting another eligible subject. 
Following their interview, each subject received 3 coupons, and no subject could be recruited more than once.  Informed consent was obtained from all participants and the study was approved by Yale University and Stellit (St. Petersburg) institutional review boards. 

\subsection{Data}

\begin{figure}
  \begin{center}
    \includegraphics[width=0.5\textwidth]{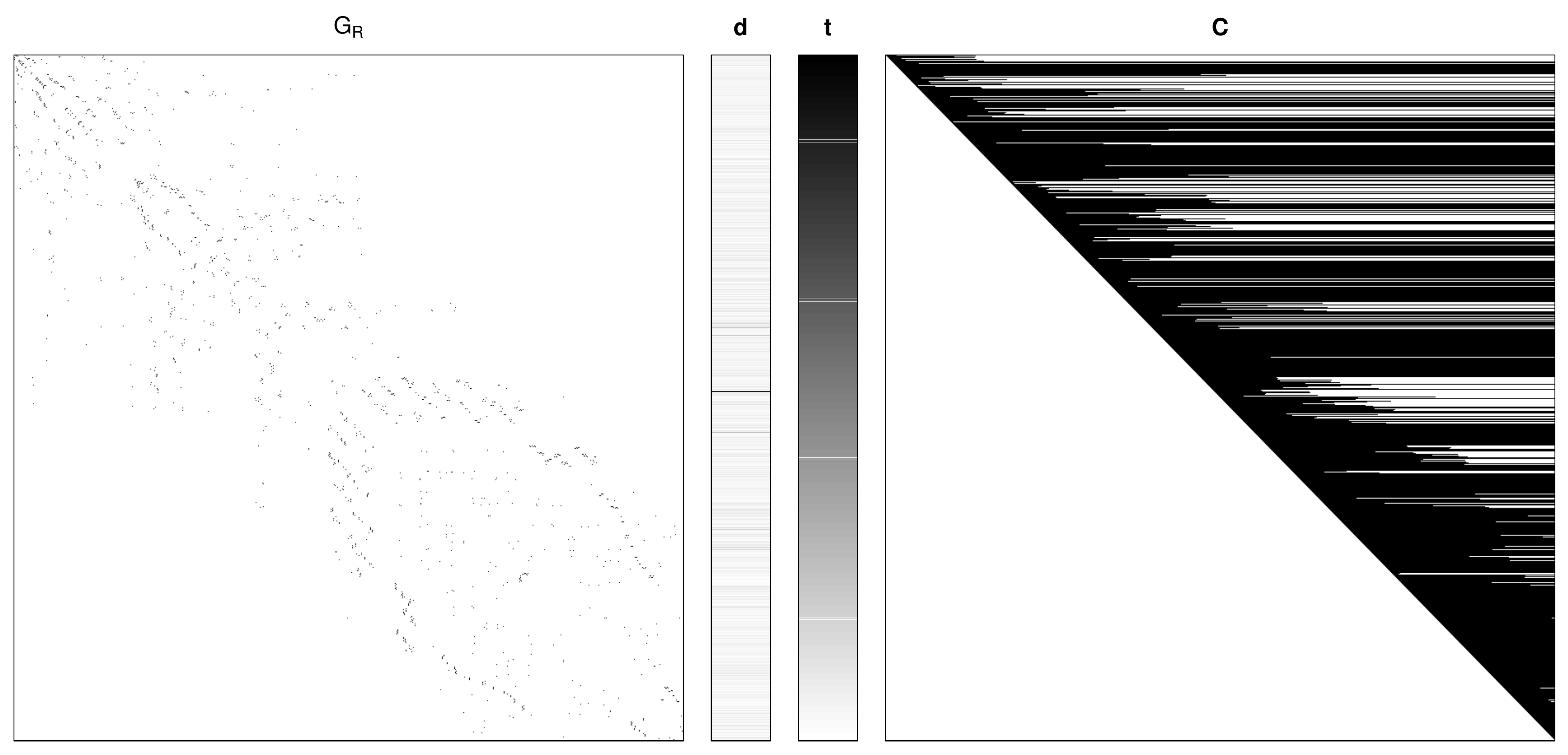}
  \end{center}
  \caption{Raw RDS data $\Y=(G_R,\bd,\mathbf{t},\C)$ extracted from study recruitment information.}
  \label{fig:obsdata}
\end{figure}

Figure \ref{fig:obsdata} shows the observed data $\Y=(G_R,\bd,\mathbf{t},\C)$ from this study.  The recruitment graph $G_R$ was constructed by matching participants' coupon ID with the IDs of coupons given to their recruiter.  The coupon matrix $\C$ was constructed by calculating the number of coupons held by each subject just before each recruitment event.  Interviews assessed network degree by asking 
\begin{quote}
  ``How many people do you know (you know one another's names) who you have seen within the last 4 weeks who inject drugs?''
\end{quote}
Define a subject's \emph{minimum degree} as the number of undirected edges incident to that subject in the recruitment graph $G_R$.  We assumed subject's network degree was accurately reported, except when a subject's reported degree was less than their minimum degree.  In these cases, we replaced the reported degree by the minimum degree.  The average reported degree of subjects was 10.3.  Interview dates and times were recorded for each subject; the elapsed time between a subject's interview and the next interview (in days) was treated as the inter-recruitment waiting time.  To more reliably estimate the edge-wise recruitment rate $\lambda$, we removed weekends and other breaks during which no interviews were scheduled.  This slightly changes the units of $\lambda$ but allows better estimation of the true waiting time distribution.  The online appendix describes the prior specification for $\lambda$.  In a few cases, the interview times for a subject and their recruit were the same, presumably because both individuals came to the interview site together.  In these cases, we resolved the tie by jittering the recruiter's interview time to be slightly earlier than the recruitee's interview time.  

Construction of $G_R$ and $\C$ revealed a minor violation of the RDS recruitment specification that merits mention: we found seven recruits whose coupon ID matched the ID of an already-redeemed coupon.  The financial reward for recruiting another eligible subject may provide a strong incentive for participants to fraudulently inflate the number of coupons they hold by creating a facsimile of the original coupon and giving it to another potential subject to redeem.  This appears to be what happened: the recruiter photocopied the original coupon, this reproduction was not detected by the interviewer, and both the new recruit and recruiter received their corresponding rewards. Rather than breaking the recruitment chain by omitting data from the seven subjects with duplicated coupon IDs, we instead artificially increase the number of coupons held by the apparent recruiter to be equal to the number of subjects who redeemed coupons bearing the ID of the recruiter.

\subsection{Results}

Overall recruitment of participants in this study was rapid: the mean time between interviews was $0.28\pm 0.74$ days.  However, the mean time between a particular subject's interview and their recruiter's interview was $23.4\pm18.0$ days, indicating that the per-edge waiting times for recruited subjects were substantially longer (the maximum waiting time from interview to recruitment was 112 days).  Indeed, this calculation is conditional on the subject actually being recruited within the study time frame, so any longer waiting times are censored by the end of the study.  We evaluated the posterior mode with $\eta$ ranging from 0.1 to 10, and in every case the estimate ranged from 0.0050 to 0.0053.  
The rate of recruitment across susceptible edges is estimated to be approximately $1/\lambda=199$ days with posterior quantiles $(186,215)$, nearly as long as the study duration of 223 days.  The apparent discrepancy between the high frequency of interviews and very slow recruitment across susceptible edges is explained by the fact that researchers observe the \emph{minimum} waiting time to recruitment across all susceptible edges at each step in the recruitment process.

\begin{figure}
  \begin{center}
    \includegraphics[width=0.7\textwidth]{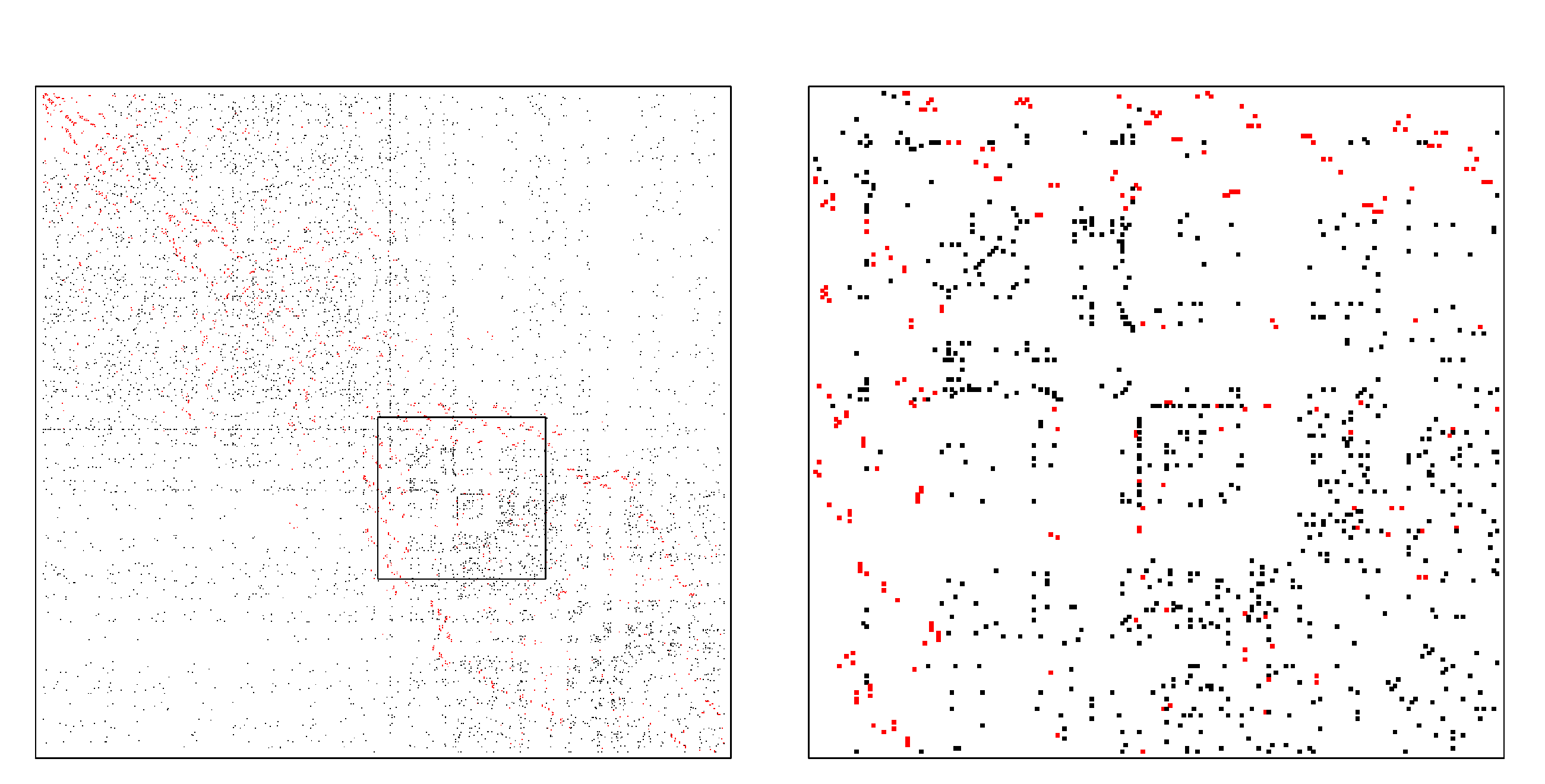} 
\end{center}
\caption{Maximum \emph{a posteriori} estimate of $G_S$ for the St. Petersburg RDS study. The left panel shows the MAP estimate of the adjacency matrix of $G_S$, and the right panel shows the inset sub-matrix in detail. Edges in the recruitment graph are shown in gray. }
  \label{fig:res}
\end{figure}

Figure \ref{fig:res} shows the MAP estimate of the adjacency matrix for all 813 sampled subjects (left) and inset detail (right). Recruitment edges appear in gray.  The apparent bands in the adjacency matrix represent high-degree individuals with many non-recruitment edges.  Probabilistic assignment of these edge ends to other recruited individuals depends on the timing of recruitments of other subjects.  The block-like structure evident in this adjacency matrix may indicate sub-networks of highly connected individuals.  Subjects recruited nearby in time may be more likely to know one another, even if they are not linked by a recruitment edge.


\section{Discussion}

\label{sec:dis}

\subsection{RDS as a network sampling method}

Nearly every paper on statistical methods for RDS data states or assumes a version of Assumption 1: the social network connecting members of the hidden population exists and determines the sampling probabilities.  But because this network is only partially observed in real-world RDS studies, Assumption 1 is usually disregarded in the formulation of statistical estimators.  Instead, researchers usually make the simplifying assumption that sampling probability is proportional to degree and does not otherwise depend on subjects' location in the network.  This simplification is justified by a thought-experiment in which the rules of the game are altered: subjects can be recruited infinitely many times, each subject receives only one coupon, and this process continues for an infinitely long time \citep{Salganik2004Sampling,Volz2008Probability,Goel2009Respondent}.  

In this paper, we have embraced Assumption 1 and its natural consequence: RDS recruitment happens across edges in the network connecting members of the hidden population.  This point of view emphasizes that RDS is more like a stochastic spreading process on a hidden network than a survey sampling method.  We define a simple continuous-time model for RDS recruitment on a hidden population graph using the kind of data obtained by every RDS study.  The model results in sensible non-uniform recruitment probabilities: the next subject is recruited with probability proportional to the number of edges they share with recruiters (Proposition 2), \emph{not their total network degree}.  Combining this model with the observed data from an RDS study allows joint estimation of the recruitment-induced subgraph $G_S$ and the waiting time parameter $\lambda$.  Most importantly, the model directly connects the observed data to the recruitment process on the underlying network.

This approach yields two computational benefits.  First, the time required to evaluate the likelihood via Proposition \ref{prop:lik} is a function of the sample size $n$ alone, and does not depend on the population size $N$, which is likely to be much larger.  In particular, we never simulate unobserved portions of the population network $G$; the ERGM \eqref{eq:lik} specifies a probability model for the recruitment-induced subgraph $G_S$ only.  In contrast, some researchers dealing with partially observed network data marginalize over the entire unsampled portion of the graph, which may be burdensome or impossible for large $N$ \citep{Gile2015Network}.  Second, the likelihood ratios in Proposition \ref{prop:ratio} do not require computation of the matrix products in \eqref{eq:s}.  Instead, \eqref{eq:addratio} and \eqref{eq:remratio} only depend on a \emph{change statistic} that can be efficiently updated. 

Our approach is unique because it uses all the available data $\Y=(G_R,\bd,\mathbf{t},\C)$ from real-world RDS studies. Several researchers have attempted to estimate the population degree distribution, but use only $G_R$ and $\bd$ (and sometimes $\bd$ alone), ignoring $\mathbf{t}$ and $\C$ \citep{Salganik2004Sampling,Volz2008Probability,Gile2011Improved,Handcock2015Estimating,Gile2015Network}.  \citet{Berchenko2013Modeling} give a formulation of recruitment event intensity similar to Assumption \ref{assump:obs} by employing a multi-type epidemic model in which active recruiters correspond to infective individuals. In their model, the rate of recruitment of a new subject with degree $k$ is proportional to the product of the number of active recruiters and the number of susceptible subjects with degree $k$.  However, they use only $\bd$, $\mathbf{t}$, and $\C$, but do not take advantage of the topological information contained in $G_R$.

\subsection{RDS and missing data}

Historically, there have been two major statistical objections to RDS as a survey design for inference of population quantities.  First, sampling probabilities cannot be computed directly from the observed data, without additional assumptions \citep{Gile2010Respondent,Gile2011Improved}. Second, there may be statistical dependence between the traits of a given subject and their neighbors (particularly their recruiter) in the network \citep{Heckathorn1997Respondent,Heckathorn2002Respondent,Tomas2011Effect,Fisher2014Stickiness}.  This dependency might be due to homophily -- the tendency for people to form social ties with others similar to them -- or preferential recruitment of certain types of people, conditional on existing social ties.  Clearly, the network structure local to the seeds and recruitment chain encodes the sampling probabilities and the statistical dependencies between subjects' attributes.  This leads us to the conclusion that a fundamental obstacle to principled statistical inference for RDS is \emph{missing data}: in RDS not all network neighbors of a vertex $i$ are observed, either because they remain unsampled, or because the recruitment graph $G_R$ does not reveal a tie between $i$ and the sampled vertices to which it is connected.  Objections to RDS typically understate the information about this network contained in the recruitment graph $G_R$ and the time series of interviews.  Our results -- revealing the graphical structure of data obtained by RDS -- raise the possibility that researchers can account for both of these sources of missing data without imposing a prior parametric graph model for the hidden network.  

Although the network may be of interest for sociological reasons, it can also be viewed as a nuisance parameter when population attributes are of primary interest.  Our simulation results and the St. Petersburg application show that the data from RDS studies contain information about the topology of $G_S$.  
Marginalizing (integrating) over the recruitment-induced subgraph $G_S$ can be understood as multiple imputation, repeatedly filling in the missing data in accordance with its distribution under the model \citep{Little1986Statistical,Huisman2009Imputation,Koskinen2010Analysing,Koskinen2013Bayesian}.  In the absence of any other information, we could marginalize over compatible graphs in $\mathcal{C}(G_R,\mathbf{d})$ with respect to the uniform distribution.  However, the reconstructed graph would be subject to two types of reconstruction inaccuracy.  First, for three sampled vertices $i$, $j$, and $k$ with at least one pendant edge each, the uniform distribution provides no basis to distinguish an edge $\{i,j\}$ from an edge $\{i,k\}$ unless a recruitment event took place along one of those edges.  For any given pendant edge, there are usually many more incorrect ways to connect it to sampled vertices than there are correct ways.  Second, marginalization with respect to the uniform distribution usually results in inclusion of too many or too few edges overall in $G_S$.  To illustrate, note that the compatibility conditions in Definition \ref{def:compatibility} provide the bounds $n-|M| \leq |E_S| \leq \frac{1}{2}{\sum_{i=1}^n} d_i$. But the uniform distribution over subgraphs in the set $\mathcal{C}(G_R,\mathbf{d})$ does not result in the uniform distribution over the set $\{ n-|M|, \ldots, \frac{1}{2}{\sum_{i=1}^n} d_i\}$.  Usually there are many more graphs $\widehat{G}_S\in\mathcal{C}(G_R,\mathbf{d})$ with $|\widehat{E}_S|>|E_S|$.  The waiting time model developed in this paper provides a coherent basis for adding edges to the recruitment subgraph $G_R$, and helps ensure that estimates of $G_S$ have approximately the same number of edges as the true underlying graph. 

In conclusion, we offer a mixed message about the prospect for statistically rigorous analysis of data from real-world RDS studies.  First, current estimators for population characteristics depend on assumptions that bear little similarity to RDS recruitment processes on social networks, and do not use all the available data.  This may account for their poor performance in empirical studies.  Second, and more optimistically, data from RDS studies contain far more information about the social network connecting respondents than has been acknowledged.  Estimation of population-level characteristics should therefore proceed from knowledge about the network of sampled subjects; extrapolation to the population network requires stronger assumptions than those given in this paper.  By introducing a simple technique for probabilistic reconstruction of the recruitment-induced subgraph, we hope to offer researchers a new tool for sociological inquiry: a social network sampling method that delivers the network.


\appendix

\section{Proof of Proposition \ref{prop:unif}}

\label{app:unif}

Let $u\in R$ be a particular recruiter and let $S_u$ be the set of susceptible vertices that are neighbors of $u$ at a given time in the recruitment process.  Let $W_{ux}$ be the waiting time for $u$ to recruit its susceptible neighbor $x\in S_u$.  By Assumption \ref{assump:exp}, $W_{ux}\sim\text{Exponential}(\lambda)$ independently for each $x\in S_u$.  Given that $u$ recruits a random vertex $X$ in $S_u$ before any other recruiter, define the first recruited vertex to be 
\begin{equation}
  X = \underset{x\in S_u}{\mathrm{argmin}}\ W_{ux} .
\end{equation}
We follow the competing risks perspective of \citet[][page 188]{Lange2010Applied} and consider the joint probability
\begin{equation}
  \begin{split}
    \Pr( X=x, W_{ux}\geq t) &= \Pr(W_{ux} \geq t, W_{uk}>W_{ux}\text{ for all } k\neq x) \\
    &= \int_t^\infty \lambda e^{-\lambda s} \Pr(W_{uk} > s\text{ for all } k\neq x) \dx{s} \\
    &= \int_t^\infty \lambda e^{-\lambda s} \prod_{\substack{k\in S_u\\ k\neq x}} e^{-\lambda s} \dx{s} \\
    &= \frac{1}{|S_u|} e^{-\lambda |S_u| t}.
  \end{split}
\end{equation}
Therefore $X$ is recruited uniformly at random from $S_u$, the waiting time to this recruitment has distribution Exponential$(\lambda|S_u|)$, and $X$ and $W_{uX}$ are independent. \qed

\section{Proof of Proposition \ref{prop:wt}}

\label{app:wt}

Let $W_{u} = \min_{x\in S_u} W_{ux}$ be the waiting time to the first recruitment by recruiter $u\in R$ and let $W = \min_{u\in R} \min_{x\in S_u} W_{ux}$ be the waiting time to the first recruitment by any recruiter.  The first recruiter is $U = \mathrm{argmin}_{u\in R} W_u$ and the first recruited vertex is $X = \mathrm{argmin}_{x\in S_U} W_{Ux}$.  We again consider the joint probability of the recruited vertex $X=x$ and the waiting time $W_{ux}$,
\begin{equation}
  \begin{split}
    \Pr(X=x, W_{Ux}\geq t) &= \sum_{u\in R} \Pr(W_{ux} \geq t,\ W_{jk}>W_{ux}\text{ for all } k\in R, j\in S, \{u,x\}\neq \{j,k\})\ \indicator{u \in R_x}  \\
                           &= \sum_{u\in R_x} \int_t^\infty \lambda e^{-\lambda s} \Pr(W_{jk} > s\text{ for all } k\in R, j\in S, \{u,x\}\neq \{j,k\})\dx{s} \\
                           &= \sum_{u\in R_x} \int_t^\infty \lambda e^{-\lambda s} \prod_{j\in R} \prod_{\substack{k\in S_j\\ \{j,k\}\neq \{u,x\}}} e^{-\lambda s} \dx{s} \\
                           &= \sum_{u\in R_x} \frac{1}{\sum_{j\in R}|S_j|} \exp\left[-\lambda t \sum_{j\in R} |S_j|\right] \\
                           &= \frac{|R_x|}{\sum_{k\in S}|R_k|} \exp\left[-\lambda t \sum_{j\in R} |S_j|\right] 
  \end{split}
\end{equation}
where the last line is obtained because $\sum_{j\in R} |S_j| = \sum_{k\in S} |R_k|$.  Therefore $X\in S$ is the first recruit with probability proportional to the number of recruiters it has (equivalently, the number of susceptible edges incident to it), the waiting time $W_{UX}$ to the first recruitment is Exponential$(\lambda\sum_{j\in R}S_j)$, and $X$ and $W_{UX}$ are independent. \qed

\section{Proof of Corollary \ref{cor:recprob}}

\label{app:recprob}

Equation \eqref{eq:myrecprob} gives the probability that a given vertex $x\in S$ is recruited at the next step in the sampling process under the model described in this paper.  \citet{Gile2010Respondent} describe a recruitment process in which the recruiter $u$ is chosen first, without regard to the number of susceptible vertices linked to it. Then, conditional on the identity of the chosen $u$, a susceptible neighbors $x\in S_u$ is recruited with uniform probability $1/|S_u|$.  Marginalizing over the recruiter $u$, we find that the probability of recruiting vertex $x\in S$ in the model of \citet{Gile2010Respondent} is 
\begin{equation}
  \begin{split}
  \Pr(x\in S\text{ is recruited}) &= \sum_{u\in R} \Pr(u\text{ is recruiter}) \Pr(x\text{ is recruited}|u\text{ is recruiter}) \\
            &= \sum_{u\in R} \frac{1}{|R|} \frac{\indicator{x\in S_u}}{|S_u|} \\
            &= \frac{1}{|R|} \sum_{u\in R_x} \frac{1}{|S_u|}
  \end{split}
\end{equation}
where the last line is obtained because $x\in S_u$ if and only if $u\in R_x$.  In general, this probability distribution is not equal to \eqref{eq:myrecprob}. \qed

\section{Proof of Proposition \ref{prop:lik}}

We first give a rigorous definition of the coupon matrix $\C$. Define the function $C_i(t)$ to be 1 if subject $i$ has at least one coupon just before time $t$, and zero otherwise.  The function $C_i(t)$ is left-continuous. Let $t_j$ be the time of the $j$th recruitment event. Then define the $i,j$th element of $\C$ as 
\begin{equation}
  \C_{ij} = \lim_{t\to t_j^-} C_i(t) .
\end{equation}
where $t\to t_j-$ means that $t$ approaches $t_j$ from the left.  Recall that $\A$ is the adjacency matrix of the recruitment-induced subgraph, with rows and columns in the order of vertices' recruitment into the study.  The $i,j$th element of the matrix product $\A\C$ is the number of recruiters connected to $i$ just before the time $t_j$ of the $j$th recruitment.  Then 
\begin{equation}
  \{\A\C\}_{ij} = \sum_{k=1}^n \A_{ik} \C_{kj}
\end{equation}
is the number of possible recruiters of subject $i$ at time $t_j^-$, and 
\begin{equation}
  \sum_{i=j}^n \sum_{k=1}^n \A_{ik} \C_{kj} 
  \label{eq:ac}
\end{equation}
is the number of susceptible edges at time $t_j^-$ connecting recruiters to vertices that will eventually be sampled.  Recruiters may also have connections to vertices that are never recruited into the study. The $i$th element of the $n\times 1$ vector $\bu$ is the number of pendant edges connecting vertex $i$ to unknown/unsampled vertices.  Each of these pendant edges is susceptible while $i$ is a recruiter, so the number of susceptible pendant edges just before time $t_j$ is
\begin{equation}
  \sum_{i=j}^n \C_{ij} \bu_i .
  \label{eq:uc}
\end{equation}
Finally, the total number of susceptible edges just before time $t_j$ is the sum of \eqref{eq:ac} and \eqref{eq:uc}, 
\begin{equation}
  \mathbf{s}_j = \sum_{i=j}^n \left( \sum_{k=1}^n \A_{ik} \C_{kj}\right) + \C_{ij} \bu_i .
  \label{eq:sj}
\end{equation}
and in vector form, 
$\mathbf{s} = \text{lt}(\A\C)'\mathbf{1} + \C'\bu$ .
Now let $\w=(0,t_2-t_1,\ldots,t_n-t_{n-1})$ be the $n\times 1$ vector of waiting times between recruitments.  By Proposition \ref{prop:wt}, the random waiting time between recruitment of subject $j-1$ and $j$ has distribution $\text{Exponential}(\lambda\mathbf{s}_j)$.  For recruited vertices $j$, this waiting time $\w_j$ is fully observed and has density 
$\lambda \mathbf{s}_j \exp[-\lambda \mathbf{s}_j \w_j]$
where $j\notin M$, where $M$ is the set of seeds.  In contrast, seeds are recruited not by other vertices, but by another mechanism.  If a seed $j$ shares edges with any recruiters before it is chosen as a seed, we observe that the actual waiting time to its recruitment must be greater than the waiting time actually observed, so the density of the waiting time $\w_j$ of a seed is
  $\exp[-\lambda \mathbf{s}_j \w_j]$.
Therefore, the full likelihood of the recruitment time series is 
\begin{equation}
  \begin{split}
    L(\w|G_S,\lambda) &= \prod_{i=1}^n (\lambda \mathbf{s}_i)^{\indicator{i\notin M}} \exp[-\lambda \mathbf{s}_i\w_i ]  \\
                      &= \left( \prod_{i\notin M} \lambda \mathbf{s}_i\right) \exp[-\lambda \mathbf{s}'\w ] ,
\end{split}
\end{equation}
where $M$ is the set of seeds, as claimed. \qed 


\section{Proof of Lemma \ref{lem:updates}}

Consider the adjacency matrix $\A$ of the current estimate of the recruitment-induced subgraph $G_S$ and suppose we would like to add an edge between $i$ and $j$, where $t_i<t_j$, $\A_{ij}=\A_{ji}=0$, $\bu_i\geq 1$, and $\bu_j\geq 1$.  Define the proposal graph as $G_S^+$ with adjacency matrix $\A^+$ to be a matrix identical to $\A$, except that $\A_{ij}^+=\A_{ji}^+=1$, with $\bu^+$ identical to $\bu$ except $\bu_i^+=\bu_i-1$ and $\bu_j^+=\bu_j-1$.  By \eqref{eq:sj}, 
  \begin{equation}
    \begin{split}
  \mathbf{s}_k^+ &= \sum_{x=k}^n \left( \sum_{y=1}^n \A_{xy}^+ \C_{yk}\right) + \C_{xk} \bu_x^+ \\
  &= \sum_{x=k}^n \left( \sum_{y=1}^n (\A_{xy} + \indicator{x=i,y=j} + \indicator{x=j,y=i}) \C_{yk}\right) + \C_{xk} \bu_x \\
  &\qquad - \C_{xk} (\indicator{x=i} + \indicator{x=j}) \\
  &= \mathbf{s}_k + \indicator{i\ge k} \C_{jk} + \indicator{j\ge k}\C_{ik} - \indicator{i\ge k}\C_{ik} - \indicator{j\ge k}\C_{jk}  \\
  &= \mathbf{s}_k + 0 + (1-\indicator{j<k})\C_{ik} - (1-\indicator{i<k})\C_{ik} - (1-\indicator{j<k})\C_{jk}  \\
  &= \mathbf{s}_k - \C_{ik}\indicator{k>j} - \C_{jk} 
\end{split}
\end{equation}
where the last line is obtained because $\C_{jk}=0$ for $k<j$.  This establishes \eqref{eq:skadd}.

Now we consider removing an edge between $i$ and $j$, where $t_i<t_j$. Suppose the current estimate of the recruitment-induced subgraph is $\A$ with $\A_{ij}=\A_{ji}=1$ with no recruitment taking place across this edge, $\{i,j\} \notin E_R$.  Define the proposal graph as $G_S^-$ with adjacency matrix $\A^-$ to be a matrix identical to $\A$, except that $\A_{ij}^-=\A_{ji}^-=0$, with $\bu^-$ identical to $\bu$ except $\bu_i^-=\bu_i+1$ and $\bu_j^-=\bu_j+1$.  Then the number of susceptible edges just before the time $t_k$ of the $k$th recruitment is 
\begin{equation}
  \begin{split}
  \mathbf{s}_k^- &= \sum_{x=k}^n \left( \sum_{y=1}^n \A_{xy}^- \C_{yk}\right) + \C_{xk} \bu_x^- \\
  &= \sum_{x=k}^n \left( \sum_{y=1}^n (\A_{xy} - \indicator{x=i,y=j} - \indicator{x=j,y=i}) \C_{yk}\right) + \C_{xk} \bu_x \\
  &\qquad + \C_{xk} (\indicator{x=i} + \indicator{x=j}) \\
  &= \mathbf{s}_k - \indicator{i\ge k} \C_{jk} - \indicator{j\ge k}\C_{ik} + \indicator{i\ge k}\C_{ik} + \indicator{j\ge k}\C_{jk}  \\
  &= \mathbf{s}_k + \indicator{k>j}\C_{ik} + \C_{jk} . 
\end{split}
\end{equation}
This gives the update formula \eqref{eq:skrem}.  \qed


\section{Proof of Proposition \ref{prop:ratio}}

For the addition of an edge between $i$ and $j$, the likelihood ratio is 
\begin{equation}
  \begin{split}
  \frac{L(\w|G_S^+,\lambda)}{L(\w|G_S,\lambda)} &= \left(\prod_{k\notin M}\frac{\mathbf{s}_k^+}{\mathbf{s}_k}\right) \exp\big[-\lambda ( \mathbf{s}^+ - \mathbf{s})'\w \big] \\
  &= \left(\prod_{k\notin M}\frac{\mathbf{s}_k^+}{\mathbf{s}_k}\right) \exp\left[\lambda \sum_{k=1}^n (\C_{ik}\indicator{k>j} + \C_{jk} )\w_k \right] .
\end{split}
\end{equation}
We have 
\begin{equation}
  \sum_{k=1}^n \C_{ik}\indicator{k>j} \w_k = t_i^* - \text{min}\{t_i^*,t_j\}
\end{equation}
and 
\begin{equation}
\sum_{k=1}^n \C_{jk} \w_k = t_j^* - t_j. 
\end{equation}
Then the ratio becomes
\begin{equation}
\frac{L(\w|G_S^+,\lambda)}{L(\w|G_S,\lambda)} = \left(\prod_{k\notin M}\frac{\mathbf{s}_k^+}{\mathbf{s}_k}\right) \exp\left[\lambda (t_i^* - \text{min}\{t_j,t_i^*\} + t_j^* - t_j)\right] .
 \end{equation}
For the removal of an edge between $i$ and $j$, the same arguments apply.  The likelihood ratio is 
\begin{equation}
  \begin{split}
  \frac{L(\w|G_S^-,\lambda)}{L(\w|G_S,\lambda)} &= \left( \prod_{k\notin M} \frac{\mathbf{s}_k^-}{\mathbf{s}_k} \right) \exp\big[-\lambda ( \mathbf{s}^- - \mathbf{s})'\w \big] \\ 
                                                &= \left(\prod_{k\notin M}\frac{\mathbf{s}_k^-}{\mathbf{s}_k}\right) \exp\left[-\lambda \sum_{k=1}^n (\C_{ik}\indicator{k>j} + \C_{jk})\w_k \right] \\
 &= \left(\prod_{k\notin M}\frac{\mathbf{s}_k^-}{\mathbf{s}_k}\right) \exp\left[-\lambda (t_i^* - \text{min}\{t_j,t_i^*\} + t_j^* - t_j)\right] ,
\end{split}
\end{equation}
as claimed.  \qed


\section{Simulation results} 

Let $\hat{\A}$ be the adjacency matrix of the estimated subgraph $\widehat{G}_S$ and let $\A$ be the adjacency matrix of the true subgraph $G_S$. We measure the accuracy, true positive rate (TPR), and true negative rate (TNR) of each estimated subgraph.  These measures are defined as follows:
\begin{equation}
  \begin{split}
    \text{Accuracy}(\hat{\A}, \A) &= \left. \sum_{i<j} \indicator{\hat{\A}_{ij} = \A_{ij}} \middle/ \binom{n}{2}\right.  \\
         \text{TPR}(\hat{\A}, \A) &= \left. \sum_{i<j} \indicator{\hat{\A}_{ij}=1 \text{ and } \A_{ij}=1} \middle/\sum_{i<j} \indicator{\hat{\A}_{ij} = 1} \right. \\
         \text{TNR}(\hat{\A}, \A) &= \left. \sum_{i<j} \indicator{\hat{\A}_{ij}=0 \text{ and } \A_{ij}=0} \middle/ \sum_{i<j} \indicator{\hat{\A}_{ij} = 0}\right.  .
\end{split}
\end{equation}

In every simulation, we generate an RDS sample of $n=500$ subjects, starting from $|M|=10$ seeds with three coupons per recruit.  Since the choice of time scale is arbitrary, we set $\lambda=1$ for every simulation.  We first evaluate reconstruction accuracy by simulating RDS on random undirected networks $G=(V,E)$ generated according to the \er random graph model with with total population size $N=1000$, 5000, and 10000 vertices and densities $p=5/N$, $10/N$, and $15/N$.  In addition, we evaluate the performance of reconstruction on a real-world network: Project 90 surveyed networks of heterosexuals at high risk of contracting HIV in Colorado Springs, CO, USA from 1988-1990 \citep{woodhouse1994mapping,klovdahl1994social,Rothenberg1995Social,Potterat2004Network}. The network data $G$ in the Project 90 data consist of $|V|=5492$ individuals and $|E|=43288$ edges. Network edges represent social, sexual, or drug use links between individuals. The Project 90 data have been used in other simulation studies to evaluate the performance of RDS estimators \citep{Goel2010Assessing}.

Table \ref{tab:sim} shows estimate summaries; each row aggregates the results of 100 simulations on distinct networks. Conditional on each simulated network, we simulate the recruitment process and report the mean and SD of estimated quantities over the 100 repetitions.  We report the parameters $N$ and $p$ used in the network simulation, the prior standard deviation (SD) of $\lambda$, the mean and standard deviation (SD) of accuracy, TPR, and TNR for reconstruction of $G_S$, and the mean and SD of estimates of $\lambda$.  Accuracy and TNR are generally very high, with lower values of TPR.  The high values of accuracy and TNR indicate that the reconstruction method recovers the true density of $G_S$ fairly well on average.  Assignment of the non-recruitment edges is more difficult, and TPR is lower.  Figure \ref{fig:simexample} shows an example in which the general structure of the adjacency matrix are recovered, but individual edges (shown as black entries in the adjacency matrix) may not always be correctly placed.  The overall accuracy of edge inference depends on the pattern of coupon use and the structure of the recruitment graph.  Accuracy is strongly affected by the proportion of recruitment edges in $G_S$: $G_R$ is always a subgraph of $G_S$, so these edges are always present in estimates of $G_S$. Therefore simulated datasets with low edge density contain very few non-recruitment edges in $G_S$ and hence reconstructions of $G_S$ enjoys very high accuracy and high TNR, while TPR is lower.  More dense subgraphs generally have higher TPR.  Some estimates of $\lambda$ exhibit small upward bias, which is alleviated by more a informative prior for $\lambda$.

\begin{table}
  \centering
  \begin{small}
  \setlength{\tabcolsep}{4pt}
\begin{tabular}{rrrrrrrrrrrrrrrrrrrrr}
  \hline
\multicolumn{2}{c}{Simulated} \\
\multicolumn{2}{c}{Network} && Prior && \multicolumn{2}{c}{Accuracy} && \multicolumn{2}{c}{TPR} && \multicolumn{2}{c}{TNR} && \multicolumn{2}{c}{$\lambda$} \\ \cline{1-2} \cline{4-4} \cline{6-7} \cline{9-10} \cline{12-13} \cline{15-16}
$N$ & $Np$  && SD$_\lambda$ && Mean & SD && Mean & SD && Mean & SD && Mean & SD \\ 
  \hline
  1000 & 5  && 1.00 && 0.994 & 2.9E-4 && 0.574 & 1.3E-2 && 0.997 & 2.1E-4 && 1.292 & 9.8E-2 \\ 
       &    && 0.10 && 0.994 & 3.1E-4 && 0.589 & 1.5E-2 && 0.997 & 2.4E-4 && 1.093 & 2.6E-2 \\ 
       &    && 0.01 && 0.994 & 2.8E-4 && 0.600 & 1.7E-2 && 0.997 & 2.1E-4 && 1.001 & 4.1E-4 \\ 
       & 10 && 1.00 && 0.985 & 5.4E-4 && 0.429 & 1.4E-2 && 0.994 & 5.7E-4 && 1.401 & 1.0E-1 \\ 
       &    && 0.10 && 0.986 & 4.9E-4 && 0.455 & 1.2E-2 && 0.994 & 4.8E-4 && 1.103 & 2.3E-2 \\ 
       &    && 0.01 && 0.986 & 5.5E-4 && 0.468 & 1.5E-2 && 0.994 & 3.9E-4 && 1.001 & 3.8E-4 \\ 
       & 15 && 1.00 && 0.977 & 7.2E-4 && 0.374 & 1.6E-2 && 0.991 & 8.2E-4 && 1.409 & 1.1E-1 \\ 
       &    && 0.10 && 0.979 & 7.0E-4 && 0.403 & 1.4E-2 && 0.990 & 7.5E-4 && 1.104 & 2.6E-2 \\ 
       &    && 0.01 && 0.980 & 8.3E-4 && 0.421 & 1.7E-2 && 0.990 & 6.3E-4 && 1.001 & 4.3E-4 \\ 
  5000 & 5  && 1.00 && 0.996 & 4.7E-4 && 0.542 & 3.2E-2 && 0.999 & 7.7E-5 && 1.401 & 9.7E-2 \\ 
       &    && 0.10 && 0.997 & 5.0E-4 && 0.617 & 4.8E-2 && 0.999 & 8.2E-5 && 1.141 & 2.6E-2 \\ 
       &    && 0.01 && 0.998 & 5.7E-4 && 0.717 & 7.3E-2 && 0.999 & 6.7E-5 && 1.002 & 7.4E-4 \\ 
       & 10 && 1.00 && 0.990 & 1.0E-3 && 0.339 & 2.6E-2 && 0.999 & 1.3E-4 && 1.540 & 1.1E-1 \\ 
       &    && 0.10 && 0.993 & 1.1E-3 && 0.440 & 4.8E-2 && 0.999 & 1.3E-4 && 1.149 & 2.3E-2 \\ 
       &    && 0.01 && 0.995 & 1.1E-3 && 0.537 & 8.2E-2 && 0.999 & 1.1E-4 && 1.003 & 6.1E-4 \\ 
       & 15 && 1.00 && 0.984 & 1.5E-3 && 0.258 & 2.0E-2 && 0.998 & 1.7E-4 && 1.576 & 1.4E-1 \\ 
       &    && 0.10 && 0.989 & 1.7E-3 && 0.355 & 4.5E-2 && 0.998 & 1.7E-4 && 1.150 & 2.5E-2 \\ 
       &    && 0.01 && 0.992 & 1.9E-3 && 0.472 & 9.0E-2 && 0.998 & 1.5E-4 && 1.003 & 8.0E-4 \\ 
 10000 & 5  && 1.00 && 0.996 & 4.4E-4 && 0.546 & 3.2E-2 && 1.000 & 5.5E-5 && 1.434 & 1.1E-1 \\ 
       &    && 0.10 && 0.997 & 5.2E-4 && 0.635 & 5.0E-2 && 1.000 & 5.4E-5 && 1.146 & 2.9E-2 \\ 
       &    && 0.01 && 0.998 & 5.9E-4 && 0.727 & 7.7E-2 && 1.000 & 5.7E-5 && 1.003 & 6.7E-4 \\ 
       & 10 && 1.00 && 0.991 & 1.2E-3 && 0.337 & 3.8E-2 && 0.999 & 7.7E-5 && 1.541 & 1.4E-1 \\ 
       &    && 0.10 && 0.994 & 1.0E-3 && 0.441 & 4.7E-2 && 0.999 & 7.9E-5 && 1.158 & 2.6E-2 \\ 
       &    && 0.01 && 0.996 & 1.2E-3 && 0.569 & 9.4E-2 && 0.999 & 7.8E-5 && 1.003 & 7.3E-4 \\ 
       & 15 && 1.00 && 0.986 & 1.6E-3 && 0.246 & 2.3E-2 && 0.999 & 1.1E-4 && 1.578 & 1.1E-1 \\ 
       &    && 0.10 && 0.991 & 1.8E-3 && 0.354 & 5.7E-2 && 0.999 & 1.0E-4 && 1.161 & 2.7E-2 \\ 
       &    && 0.01 && 0.994 & 1.8E-3 && 0.481 & 1.1E-1 && 0.999 & 9.0E-5 && 1.003 & 7.4E-4 \\ 
   \hline
 \multicolumn{2}{c}{Project 90} && 1.00 && 0.973 & 1.6E-3 && 0.376 & 2.6E-2 && 0.989 & 1.1E-3 && 1.263 & 1.0E-1 \\
       &    && 0.10 && 0.974 & 1.3E-3 && 0.370 & 2.5E-2 && 0.988 & 9.5E-4 && 1.085 & 3.6E-2 \\ 
       &    && 0.01 && 0.974 & 1.5E-3 && 0.374 & 2.3E-2 && 0.987 & 9.3E-4 && 1.001 & 4.1E-4 \\ 
   \hline
\end{tabular}
\end{small}

\caption{Simulation results for RDS on \er random networks and the Project 90 network. The recruitment rate $\lambda$ in every simulation is 1. Each row summarizes 100 simulations of RDS on different random networks under the given network parameters $N$ and $p$.  The prior SD of $\lambda$ is given in the next column. The mean and SD of accuracy, TPR, and TNR for reconstruction of $G_S$ and the mean MAP estimate and SD of $\lambda$ are in the last three columns.  Below, simulation results for RDS on the Project 90 network are given. }
  \label{tab:sim}
\end{table}

\begin{table}
  \centering
    \setlength{\tabcolsep}{4pt}
\begin{tabular}{rrrrrrrrrrrrrrrrrrrrr}
 \hline
 && Prior && \multicolumn{2}{c}{Accuracy} && \multicolumn{2}{c}{TPR} && \multicolumn{2}{c}{TNR} && \multicolumn{2}{c}{$\lambda$} \\ \cline{3-3} \cline{5-6} \cline{8-9} \cline{11-12} \cline{14-15}
$\delta$  && SD$_\lambda$ && Mean & SD && Mean & SD && Mean & SD && Mean & SD \\ 
  \hline
  0.5 && 1.00 && 0.974 & 1.6E-3 && 0.394 & 2.9E-2 && 0.990 & 1.0E-3 && 19.816 & 3.4E-0 \\ 
      && 0.10 && 0.986 & 1.1E-3 && 0.969 & 6.0E-2 && 0.986 & 9.1E-4 && 3.872 & 2.3E-1 \\ 
      && 0.01 && 0.986 & 9.3E-4 && 1.000 & 3.3E-4 && 0.986 & 9.4E-4 && 1.043 & 8.8E-4 \\ 
  0.6 && 1.00 && 0.974 & 1.5E-3 && 0.387 & 3.0E-2 && 0.990 & 1.0E-3 && 8.748 & 1.4E-0  \\ 
      && 0.10 && 0.985 & 1.9E-3 && 0.814 & 1.2E-1 && 0.986 & 9.2E-4 && 2.698 & 1.7E-1 \\ 
      && 0.01 && 0.986 & 9.6E-4 && 0.998 & 8.2E-3 && 0.986 & 9.6E-4 && 1.037 & 1.3E-3 \\ 
  0.7 && 1.00 && 0.973 & 1.5E-3 && 0.385 & 2.6E-2 && 0.989 & 1.0E-3 && 4.390 & 5.9E-1 \\ 
      && 0.10 && 0.981 & 2.2E-3 && 0.596 & 9.1E-2 && 0.986 & 1.0E-3 && 1.885 & 9.4E-2 \\ 
      && 0.01 && 0.986 & 1.1E-3 && 0.977 & 4.6E-2 && 0.986 & 1.0E-3 && 1.026 & 2.4E-3 \\ 
  0.8 && 1.00 && 0.973 & 1.4E-3 && 0.383 & 2.5E-2 && 0.989 & 9.4E-4 && 2.710 & 2.6E-1 \\ 
      && 0.10 && 0.978 & 2.0E-3 && 0.464 & 4.3E-2 && 0.987 & 1.0E-3 && 1.441 & 7.7E-2 \\ 
      && 0.01 && 0.984 & 2.3E-3 && 0.794 & 1.4E-1 && 0.986 & 1.1E-3 && 1.013 & 2.6E-3 \\ 
  0.9 && 1.00 && 0.974 & 1.6E-3 && 0.380 & 2.5E-2 && 0.989 & 1.0E-3 && 1.736 & 1.6E-1 \\ 
      && 0.10 && 0.975 & 1.6E-3 && 0.399 & 2.3E-2 && 0.987 & 9.3E-4 && 1.197 & 3.9E-2 \\ 
      && 0.01 && 0.978 & 2.3E-3 && 0.458 & 6.2E-2 && 0.987 & 1.0E-3 && 1.004 & 1.1E-3 \\ 
\hline
1.1 && 1.00 && 0.974 & 1.7E-3 && 0.379 & 2.7E-2 && 0.989 & 1.1E-3 && 0.929 & 7.2E-2 \\ 
    && 0.10 && 0.973 & 1.5E-3 && 0.380 & 2.7E-2 && 0.989 & 1.0E-3 && 0.970 & 3.6E-2 \\ 
    && 0.01 && 0.973 & 1.6E-3 && 0.374 & 2.5E-2 && 0.989 & 1.0E-3 && 0.999 & 8.2E-4 \\ 
1.2 && 1.00 && 0.973 & 1.5E-3 && 0.373 & 2.4E-2 && 0.988 & 1.0E-3 && 0.731 & 5.3E-2 \\ 
    && 0.10 && 0.974 & 1.7E-3 && 0.394 & 2.8E-2 && 0.990 & 9.4E-4 && 0.834 & 3.3E-2 \\ 
    && 0.01 && 0.974 & 2.0E-3 && 0.405 & 3.3E-2 && 0.991 & 1.0E-3 && 0.995 & 1.6E-3 \\ 
1.3 && 1.00 && 0.973 & 1.5E-3 && 0.370 & 2.3E-2 && 0.988 & 1.2E-3 && 0.587 & 4.4E-2 \\ 
    && 0.10 && 0.974 & 1.9E-3 && 0.401 & 3.2E-2 && 0.990 & 1.0E-3 && 0.724 & 3.0E-2 \\ 
    && 0.01 && 0.976 & 2.1E-3 && 0.436 & 3.4E-2 && 0.993 & 1.0E-3 && 0.987 & 2.4E-3 \\ 
1.4 && 1.00 && 0.973 & 1.5E-3 && 0.366 & 2.4E-2 && 0.988 & 1.1E-3 && 0.488 & 3.6E-2 \\ 
    && 0.10 && 0.974 & 1.8E-3 && 0.404 & 2.9E-2 && 0.990 & 1.0E-3 && 0.626 & 2.4E-2 \\ 
    && 0.01 && 0.977 & 1.9E-3 && 0.460 & 3.3E-2 && 0.994 & 9.8E-4 && 0.978 & 3.1E-3 \\ 
1.5 && 1.00 && 0.973 & 1.6E-3 && 0.352 & 2.2E-2 && 0.988 & 1.1E-3 && 0.412 & 3.0E-2 \\ 
    && 0.10 && 0.974 & 2.1E-3 && 0.400 & 3.4E-2 && 0.990 & 1.0E-3 && 0.542 & 2.1E-2 \\ 
    && 0.01 && 0.978 & 2.1E-3 && 0.481 & 3.6E-2 && 0.995 & 9.1E-4 && 0.968 & 3.4E-3 \\ 
   \hline
\end{tabular}
\caption{Simulation results for RDS on the Project 90 data under mis-specification of the waiting time distribution.  The mean waiting time to recruitment is distributed as Gamma$(\delta,\delta)$, for the given values of $\delta$.  When $\delta=1$, the waiting time distribution is correctly specified; these results are given in the last three lines of Table \ref{tab:sim}. }
  \label{tab:simdelta}
\end{table}

In real-world RDS studies, Assumption \ref{assump:exp} may be violated.  It is therefore important to assess the performance of subgraph reconstruction and estimation of $\lambda$ when the waiting time distribution is mis-specified and Assumption \ref{assump:exp} does not hold. To this end, we draw random edge-wise waiting times $T_{ij}$ from the Gamma distribution with density $f(t) = \lambda^\delta t^{\delta-1} e^{-\delta t}/ \Gamma(\delta)$ with shape and rate parameters $\delta>0$.  Setting $\delta=1$ recovers the Exponential$(\lambda=1)$ distribution.  When $0<\delta<1$, the Gamma density decays monotonically with the waiting time.  When $\delta>1$, the density has a nonzero mode. In simulations under this waiting time distribution, we fix the mean waiting time at $\E[T_{ij}]=1$ and vary $\delta$.  In this way, $\delta$ provides a conventient continuous parameter to change the magnitude of mis-specification of the waiting time model.  

Table \ref{tab:simdelta} gives the results for Gamma-distributed waiting times, where $\delta$ is specified.  The prior SD of $\lambda$, accuracy, TPR, TNR, and the mean and SD of estimates of $\lambda$ are given.  In general, accuracy, TPR, and TNR are roughly the same as in the Exponential($\lambda=1$) case. But as expected, estimates of $\lambda$ under a mis-specified waiting time distribution appear to be subject to bias.  When $\delta<1$, $\lambda$ is typically over-estimated; when $\delta>1$, $\lambda$ is usually under-estimated.  We can explain the relative robustness of reconstruction by recalling two features of the proposed framework.  First, the compatibility conditions (Definition \ref{def:compatibility}) impose strong constraints on the topology of $G_S$ when $G_R$ and $\bd$ are observed.  These constraints are in effect regardless of whether the waiting time model is correctly specified, and serve to ensure that all edges in $G_R$ are correctly estimated. Second, under any model in which the recruitment times across edges are independent, the rate of new recruitments is positively associated with the number of susceptible edges.  Under the exponential model this relationship is linear, and under other waiting time models the relationship may be non-linear (and in general depends on time waited along each susceptible edge up to the current time).  

\section{Prior for $\lambda$ in the St. Petersburg application}

To obtain a sensible prior distribution for the edge-wise recruitment rate $\lambda$, we adopt an empirical Bayesian approach based on bounding the maximum likelihood estimate of $\lambda$ given by \eqref{eq:lamhat}.  First, observe that the maximum number of susceptible edges that can be added by a vertex $i$ with degree $d_i$ is $d_i$ minus one if $i$ was recruited by another subject, $d_i-\indicator{i\notin M}$.  Then the maximum number of susceptible edges at step $k$ in the recruitment process is $\mathbf{s}_k = \sum_{i=1}^{k-1} d_i - \indicator{i\notin M}$.  Therefore a minimum estimate of $\lambda$ is 
\[ \lambda_\text{lo} = \frac{n-|M|}{ \sum_{k=1}^n \w_i \sum_{i=1}^{k-1} (d_i - \indicator{i\notin M}) } . \]
Now let $n_i$ be the number of subjects recruited by subject $i$.  The minimum number of susceptible edges at step $k$ is $\mathbf{s}_k = \sum_{i=1}^{k-1} n_i - \indicator{i\notin M}$, and the maximum estimate of $\lambda$ is 
\[ \lambda_\text{hi} = \frac{n-|M|}{ \sum_{k=1}^n \w_i \sum_{i=1}^{k-1} (n_i - \indicator{i\notin M}) } . \]
Applying these bounds to the St. Petersburg data yields $\lambda_\text{lo}=9.8\times 10^{-4}$ and $\lambda_\text{hi}=4.2\times 10^{-2}$.  We therefore specify a prior distribution for $\lambda$ that takes most of its mass in the interval $[\lambda_\text{lo},\lambda_\text{hi}]$.  Let $\lambda_\text{mean}=(\lambda_\text{lo}+\lambda_\text{hi})/2=0.022$. Suppose $\eta>0$ is given and let $\xi = \eta/\lambda_\text{mean}$.  Now by varying $\eta$, we obtain a family of Gamma prior distributions with mean $\lambda_\text{mean}$.


\bibliographystyle{asr}
\bibliography{rds}

\end{document}